\pdfoutput=1

\documentclass[11pt,twoside,a4paper,cmspaper,final,collab]{cms-tdr}
\def\svnVersion{454512P}\def\cmsCernNoTag{CERN-EP-2017-234}\def\cmsCernDate{\today}\def\cmsMessage{Published in the Journal of High Energy Physics as \href{http://dx.doi.org/10.1007/JHEP03(2018)172}{\doi{10.1007/JHEP03(2018)172.}}}

\begin{document}\cmsNoteHeader{SMP-17-002}

\hyphenation{had-ron-i-za-tion}
\hyphenation{cal-or-i-me-ter}
\hyphenation{de-vices}
\RCS$Revision: 446679 $
\RCS$HeadURL: svn+ssh://svn.cern.ch/reps/tdr2/papers/SMP-17-002/trunk/SMP-17-002.tex $
\RCS$Id: SMP-17-002.tex 446679 2018-02-18 13:58:57Z alverson $

 \newlength\cmsFigWidth
\ifthenelse{\boolean{cms@external}}{\setlength\cmsFigWidth{0.85\columnwidth}}{\setlength\cmsFigWidth{0.4\textwidth}}
\ifthenelse{\boolean{cms@external}}{\providecommand{\cmsLeft}{top\xspace}}{\providecommand{\cmsLeft}{left}}
\ifthenelse{\boolean{cms@external}}{\providecommand{\cmsRight}{bottom\xspace}}{\providecommand{\cmsRight}{right}}
 \cmsNoteHeader{SMP-17-002}

\newcommand{\qT}  {\ensuremath{q_{\mathrm{T}}}\xspace}
\newcommand{\Mll} {\ensuremath{m_{\ell\ell}}\xspace}
\providecommand{\PYTHIASIX} {\PYTHIA{}6\xspace}
\providecommand{\PYTHIAEIGHT} {\PYTHIA{}8\xspace}
\providecommand{\RESBOS} {\textsc{ResBos}\xspace}
\providecommand{\AMCATNLO}{\MADGRAPH{}5\_a\MCATNLO\xspace}
\providecommand{\aMCATNLO} {a\MCATNLO}

\title{Measurement of differential cross sections in the kinematic\\ angular variable $\phi^*$ for inclusive $\cPZ$ boson production\\ in pp collisions at $\sqrt{s} = 8\TeV$}

 \date{\today}

 \abstract{Measurements of differential cross sections $\rd\sigma / \rd\phi^*$  and double-differential cross sections $\rd^2\sigma / \rd\phi^*\rd\abs{y}$  for inclusive Z~boson production are presented using the dielectron and dimuon final states. The kinematic observable $\phi^*$  correlates with the dilepton transverse momentum but has better resolution, and $y$ is the dilepton rapidity. The analysis is  based on  data  collected  with the CMS experiment at a centre-of-mass energy  of 8\TeV corresponding to an integrated luminosity of 19.7\fbinv.   The normalised cross section $(1 / \sigma)\,\rd\sigma / \rd\phi^*$, within the fiducial kinematic region,  is measured with a precision of better than 0.5\% for $\phi^* < 1$. The measurements  are compared to theoretical predictions and they agree, typically,  within few percent.
 }

\hypersetup{%
pdfauthor={CMS Collaboration},%
pdftitle={Measurement of differential cross sections in  the phi* variable for inclusive Z boson production in pp collisions at sqrt(s)= 8 TeV},%
pdfsubject={CMS},%
pdfkeywords={CMS, physics, Drell--Yan, phistar}}

\maketitle

 \section{Introduction}
 \label{sec:introduction}

The neutral current Drell--Yan (DY) process, $\qqbar \to \cPZ/\gamma^{*} \to \ell^+\ell^-$, where $\ell$  is either an electron or a muon, is one of the best studied physics processes at the CERN LHC. The total and differential cross sections have been calculated theoretically at next-to-next-to-leading order (NNLO) accuracy in the strong coupling $\alpS$~\cite{Anastasiou:2003ds,Melnikov:2006kv,Li:2012wna,Alioli:2015toa}. The differential cross section as a function of dilepton invariant mass $\ddinline{\sigma}{m_{\ell\ell}}$ has been measured by the LHC experiments at different centre-of-mass energies~\cite{Chatrchyan:2013tia,CMS:2014jea,Aad:2014qja, Aad:2013iua}. Theoretical calculations reproduce the measurements over nine orders of magnitude at the level of a few percent.

The large production cross section and the experimentally clean final state of the DY process allow for detailed studies of kinematic distributions that serve as stringent tests of the perturbative calculations. One of the most interesting observables is the transverse momentum $\qT$ of the \cPZ{} boson, which is related to its production mechanism. The lower range of $\qT$ values are caused by multiple soft-gluon emissions, whereas high $\qT$ values result from the emission of one or more hard partons in association with the  \cPZ{} boson. Another interesting observable is the rapidity $y$ of the \cPZ{} boson which depends on the difference in momentum between the parent partons in the colliding protons; therefore, the cross section as a function of $y$ depends on the parton distribution functions (PDF). The  $\qT$ spectrum of the \cPZ{} boson has been measured by the  ATLAS, CMS and LHCb Collaborations at $\sqrt{s} = 7$\TeV~\cite{Aad:2011gj,Chatrchyan:2011wt,Aaij:2012mda}.  Recently, both the CMS and ATLAS Collaborations have extended the study at 8\TeV by performing double-differential measurements as functions of $\qT$ and $y$~\cite{Khachatryan:2015oaa,Aad:2015auj}.  Calculations based on fixed-order perturbative quantum chromodynamics (QCD)~\cite{Hoeche:2014aia} describe these measurements fairly well.

A thorough understanding of the $\qT$ spectra of the electroweak vector bosons  is essential for high-precision measurements at the LHC, in particular that of the mass of the $\PW$~boson. Furthermore, the theoretical calculation of the transverse momentum distribution for the Higgs boson produced in gluon-gluon fusion at the LHC involves Sudakov form factors~\cite{Collins:1989bt}, which are closely related to those appearing in the calculations for $\qT$. Thus precise measurements of vector boson production  are important for validating the theoretical calculations of Higgs boson production at the LHC.

An important issue in the accurate measurement of the differential cross section $\ddinline{\sigma}{\qT}$ is the experimental resolution of $\qT$, which is dominated by the uncertainties in the magnitude of the transverse momenta of the leptons from the decay of the \cPZ{} boson. The angles subtended by the leptons, however, are measured more precisely due to the excellent spatial resolution of the CMS tracking system. A kinematic quantity $\phi^*$~\cite{Banfi:2010cf,Banfi:2012du,Marzani:2013nza}, derived from these angles, is defined by the expression

\begin{equation}
\label{eq0}
\phi^*  = \tan \left( \frac{\pi -\Delta\phi}{2} \right) \, \sin(\theta^*_\eta) .
\end{equation}

The variable  $\Delta\phi$ is the opening angle between the leptons in the plane transverse to the beam axis. The variable $\theta^{*}_\eta$ indicates the scattering angle of the dileptons with respect to the beam in the boosted frame where the leptons are aligned. It is related to the pseudorapidities of the oppositely charged leptons by the relation $ \cos(\theta^*_\eta) = \tanh[ \Delta\eta / 2 ] $, where $\Delta\eta$ is the difference in pseudorapidity between the two leptons.  By construction, $\phi^*$ is greater than zero. Since $\phi^*$ depends on angular variables, the resolution of~$\phi^*$ is significantly better than that of $\qT$, especially at low $\qT$ values. Since $\phi^* \sim \qT / m_{\ell\ell}$, the range $\phi^* \le 1$ corresponds to $\qT$ up to about 100\GeV for a dilepton mass close to the nominal \cPZ{} boson mass.

The cross sections for the DY process  as a function of $\phi^*$ have been measured by the \DZERO~Collaboration at the Tevatron for $\Pp\bar\Pp$ collisions at $\sqrt{s} = 1.96\TeV$~\cite{Abazov:2010mk} and at the LHC by the ATLAS~Collaboration for $\Pp\Pp$ collisions at  7 and~$8\TeV$~\cite{Aad:2012wfa,Aad:2015auj}. In this paper, the  measurements of the differential cross section $\ddinline{\sigma}{\phi^*}$  and the  double-differential cross section ${\rd}^2 \sigma / {\rd} \phi^* {\rd}\abs{y}$ in CMS at $\sqrt{s} = 8\TeV$ are presented using data corresponding to an integrated luminosity of $\Lumi = 19.7 \pm 0.5\fbinv$.

The paper is organized as follows. A brief description of the CMS detector is
presented in Section~\ref{sec:CMSDetector}. The general features of event
reconstruction and selection for the analysis are
discussed in Section~\ref{sec:EvSelCriteria}. The details of simulated samples used
to guide and validate the measurements are given in Section~\ref{sec:datamc_description}.
Section~\ref{sec:method} states the precise definitions of the fiducial region and
the differential cross sections. Section~\ref{sec:bkgnd} describes the background
subtraction, and Section~\ref{sec:unfolding} describes how the signal distributions
are unfolded to remove the impact of resolution in the experimental measurement.
Section~\ref{sec:systematics} provides a discussion of the systematic uncertainties.
Section~\ref{sec:predictions} discusses the theoretical predictions that are
compared to the measured cross sections. Finally the results are reported and discussed
in Section~\ref{sec:results}, with a summary presented in Section~\ref{sec:conclusion}.

 \section{The CMS detector}
 \label{sec:CMSDetector}
The central feature of the CMS apparatus is a superconducting solenoid of 6\unit{m} internal diameter, providing a magnetic field of 3.8\unit{T}. Within the solenoid volume are a silicon pixel and strip tracker, a lead tungstate crystal electromagnetic calorimeter (ECAL), and a brass and scintillator hadron calorimeter (HCAL), each composed of a barrel and two endcap sections. The steel and quartz-fibre Cherenkov hadron forward calorimeters extend the pseudorapidity coverage provided by the barrel and endcap detectors. Muons are measured in the gas-ionization detectors embedded in the steel flux-return yoke outside the solenoid, with detection planes made using three technologies: drift tubes, cathode strip chambers, and resistive-plate chambers. A more detailed description of the CMS detector, together with a definition of the coordinate system used and the relevant kinematic variables, can be found in Ref.~\cite{Chatrchyan:2008zzk}.

 \section{Event reconstruction and selection}
 \label{sec:EvSelCriteria}
 Events of interest are selected using a two-tiered trigger system~\cite{Khachatryan:2016bia}. The first level (L1), composed of custom hardware processors, uses information from the calorimeters and muon detectors to select events at a rate of around 100\unit{kHz} within a time interval of less than 4\mus. The second level, known as the high-level trigger (HLT), consists of a farm of processors running a version of the full event reconstruction software optimized for fast processing, and reduces the event rate to around 1\unit{kHz} before data storage.
The events for this analysis are triggered by the presence of at least one electron  with transverse momentum $\pt > 27\GeV$ and $\abs{\eta} < 2.5$,  or at least one muon with  $\pt > 24\GeV$  and $\abs{\eta} < 2.1$. Both electrons and muons must satisfy relatively loose isolation and identification requirements compared to the off-line selection. For this analysis, the overall performance of this trigger is found to be better than the inclusive dilepton trigger.

Because of the high instantaneous luminosity, there are multiple $\Pp\Pp$ collisions within the same bunch crossing leading to  event pileup  in the detector. The average number of pileup in a triggered event during the 2012 data taking period is about 21. The reconstructed vertex with the largest value of summed physics-object $\pt^2$ is taken to be the primary $\Pp\Pp$ interaction vertex. The physics objects are the objects returned by a jet finding algorithm~\cite{Cacciari:2008gp,Cacciari:2011ma} applied to all charged tracks associated with the vertex, plus the corresponding associated missing transverse momentum.

The off-line particle-flow event algorithm~\cite{Sirunyan:2017ulk} reconstructs and identifies individual particles with an optimised combination of information from the various elements of the CMS detector. The photon energy is obtained  directly from the ECAL measurement, corrected for zero-suppression effects. Electron identification relies on the electromagnetic shower shape and other observables based on tracker and calorimeter information~\cite{Khachatryan:2015hwa}. The  barrel-endcap transition regions of the ECAL ($1.444<\abs{\eta}<1.566)$ are excluded from the acceptance.  The energy of electrons is inferred from a combination of the electron momentum at the primary vertex as determined by the tracker, the energy of the corresponding ECAL cluster, and the energy sum of all the bremsstrahlung photons spatially compatible with originating from the electron track.

Electrons originating from photon conversions are suppressed by requiring no more than one missing tracker hit and that the final hit on the reconstructed track matches an electron cluster in the ECAL. Electron candidates are rejected if they form a pair with a nearby track that is consistent with photon conversion. To ensure that the electron is consistent with a particle originating from the primary interaction vertex, the magnitude of the transverse impact parameter of the candidate track must be less than 0.02\cm, and the longitudinal distance from the primary interaction vertex is required to be less than $0.1\cm$. The momentum resolution for electrons from $\cPZ\to\Pep\Pem$ decays ranges from 1.7\% for electrons in the barrel region to 4.5\% for electrons that begin to shower before the calorimeter in the  endcaps~\cite{Khachatryan:2015hwa}.

The transverse momentum of muons is obtained from the curvature of the muon tracks in the muon detector combined with matched tracks in the silicon tracker.  Muon candidates are selected by applying minimal requirements to the track segments in both  muon  and inner tracker systems as well as consistent with small energy deposits in the calorimeters. The track associated with each muon candidate is required to have at least one hit in the pixel detector and at least five hits in different layers of the silicon tracker. The muon candidate is required to have hits in at least two different muon stations. To reject cosmic ray muons, the magnitude of the transverse impact parameter is required to be less than 0.2\cm and the longitudinal distance from the primary interaction vertex is required to be less than 0.5\cm~\cite{Chatrchyan:2012xi}. Selected muons in the range $20 <\pt < 100\GeV$ have a relative \pt resolution of 1.3--2.0\% in the barrel ($\abs{\eta}<1.2$) and less than 6\% in the endcaps ($1.2<\abs{\eta}<2.4$)~\cite{Chatrchyan:2012xi}.

The energy of  charged hadrons is determined from a combination of their momentum measured in the tracker, and the matched ECAL and HCAL energy deposits. Subsequently, it is corrected for zero-suppression effects and for the response function of the calorimeters to hadronic showers. Finally, the energy of neutral hadrons is obtained from the corresponding corrected ECAL and HCAL energies.

Events containing at least two leptons are selected, in which one lepton, consistent with the trigger, satisfies $\pt > 30\GeV$ and $\abs{\eta} < 2.1$, while the other is required to have $\pt > 20\GeV$ and $\abs{\eta} < 2.4$.  These two leptons must have the same flavour and originate from the same primary vertex. For dimuon events, the leptons must have opposite electric charges. The probability of charge misidentification is not negligible for electrons and hence this criteria is not applied to dielectron events. Events are retained if the dilepton invariant mass falls in the range $60 < \Mll < 120\GeV$.

The leptons in the DY process are usually isolated from other particles produced in the event; hence isolation criteria are useful for rejecting non-DY events. The isolation of a lepton, $I$, is defined as the ratio of the sum of the transverse momenta of the charged and neutral hadrons as well as photons that fall within a cone of radius $\Delta R = \sqrt{\smash[b]{(\Delta\eta)^{2} + (\Delta\phi)^{2}}}$ (where $\phi$ is the azimuthal angle in radians) centered on the lepton to its $\pt$. The requirement that the reconstructed charged particle tracks originate from a common primary vertex practically eliminates the pileup contribution from charged hadrons. In the case of electrons the pileup contributions for neutral hadrons and photons are estimated on a statistical basis using the approach of jet area subtraction~\cite{Cacciari:2007fd}. For muons the corresponding subtracted quantity is computed by summing up the momenta of the charged tracks not associated with the interaction vertex and multiplying the total contribution by a factor of 0.5 to account for the relative fraction of neutral and charged particles.  The values of the cone size and relative isolation optimised for electrons (muons) are $\Delta R < 0.3 (0.4)$ with $I < 0.15 (0.12)$.

Applying the full set of selection criteria, the dielectron and dimuon data samples include approximately 4.4 and 6.7~million events, respectively.

 \section{Monte Carlo simulation}
 \label{sec:datamc_description}

Samples of simulated Monte Carlo (MC) events are used for estimating the signal efficiencies and the rates of most of the background processes. An inclusive DY  signal sample generated by the \MADGRAPH (v1.3.30) leading order (LO) matrix element generator~\cite{Alwall:2011uj} that includes up to four extra partons in the calculation, is used to estimate the efficiency and to unfold the data. The parton distribution function (PDF) set CTEQ6L1~\cite{Pumplin:2002vw} is used for the generation of this sample. Parton shower and hadronisation effects are implemented by interfacing the event generator with \PYTHIASIX (v6.4.24)~\cite{Sjostrand:2006za} along with the \kt-MLM matching scheme~\cite{Alwall:2007fs} and using the Z2* tune~\cite{Chatrchyan:2013gfi,Khachatryan:2015pea} for the underlying event.

The background due to DY $\to \TT$ production is simulated in the \MADGRAPH sample used for the signal. The decays of $\tau$ leptons are described by the \TAUOLA (v1.27)~\cite{Golonka:2003xt} package. The backgrounds due to $\ttbar$ and \PW+jets events are also generated using \MADGRAPH, while dibosons (\PW\PW, \PW\cPZ\  and \cPZ\cPZ), single top quarks ($\PQt\PW$ and $\PAQt\PW$), and muon-enriched  QCD multijet samples are generated using \PYTHIASIX. The cross sections for the simulated processes are normalised to the available state-of-the-art theoretical calculations.
For the \MADGRAPH signal as well as \PW+jets samples, the total inclusive cross sections are normalised to the values obtained from the  theoretical predictions, computed using \FEWZ (v2.0)~\cite{Gavin:2010az} with the NNPDF3.0 set of PDF~\cite{Ball:2014uwa}. \FEWZ includes QCD corrections up to NNLO  and electroweak corrections up to next-to-leading order (NLO). The \ttbar rate is normalised to the predicted  cross section with NNLO+NNLL (next-to-next-to-leading logarithm) accuracy~\cite{Czakon:2011xx}. The normalisations for single top quark and diboson samples use cross section values available at NLO accuracy~\cite{Aliev:2010zk, Kant:2014oha, Campbell:2011bn, Melia:2011tj}. For QCD multijet events the simulated sample is normalised to the LO cross section.

The generated events are passed through a CMS detector simulation based on \GEANTfour~\cite{Agostinelli:2002hh}.
Minimum bias events are superposed on each of the simulated samples to account for pileup. The number
of superposed events is dictated by the distribution of the number of reconstructed primary vertices in data,
which is a function of the instantaneous luminosity.

 \section{Analysis method}
 \label{sec:method}

The fiducial region is defined by a common set of kinematic restrictions
applied to both the dielectron and the dimuon channels: one lepton with $\pt > 30\GeV$ and $\abs{\eta} < 2.1$,  a second lepton with $\pt > 20\GeV$ and $\abs{\eta} < 2.4$, and a dilepton invariant mass $60 < \Mll < 120\GeV$. The $\phi^*$ range is restricted to a value less than  3.227 so as to keep the statistical and systematic uncertainties comparable in the relevant bin.
Leptons are defined at Born level, \ie, before bremsstrahlung or  final-state radiation of  photon ~(QED-FSR).

Differential cross sections are defined within this fiducial region.
Before the spectra are unfolded (as it will be discussed later), the absolute
differential cross section is defined by
\begin{equation}
\label{eq1}
\left[\frac {\rd \sigma}{\rd\phi^*}\right] _i =
\frac{{\mathcal N}_i - {\mathcal B}_i}{{\Lumi} \, \epsilon _i \, \Delta \phi^*_i},
\end{equation}
where ${\mathcal N}_i$, ${\mathcal B}_i$, $\epsilon _i$, and $\Delta \phi^*_i$ are
the number of selected events, the estimated number of background events, the overall  efficiency, and the width of the $i$th bin of $\phi^*$, respectively, and  ${\Lumi}$ is the total integrated luminosity.

 The normalised cross section is defined as the absolute cross section divided by   the integral over all the bins of the differential distribution: $(1 / \sigma)\,\ddinline{\sigma}{\phi^*}$. The cancellation of some of the factors leads to a reduction in uncertainty, and hence the normalised cross section is more suitable for a comparison with theoretical predictions.

The double-differential cross section is defined similarly by taking into account  the width of the rapidity bin $\Delta\abs{y}_j$, and the efficiency,  defined suitably,
\begin{equation}
\label{eq2}
\left[\frac {\rd^2 \sigma}{\rd\phi^*~\rd \abs{y}}\right] _{ij}
= \frac {{\mathcal N}_{ij} - {\mathcal B}_{ij}}{{\Lumi} \, \epsilon_{ij} \, \Delta \phi^*_i \, \Delta\abs{y}_j}.
\end{equation}
The normalised double-differential cross section is given by
$(1 / \sigma) \, \rd^2 \sigma / {\rd} \phi^* {\rd}\abs{y}$.

The efficiencies for the trigger, reconstruction, identification, and isolation requirements on the leptons are obtained in bins of $\pt$ and $\abs{\eta}$ using ``tag-and-probe'' techniques~\cite{CMS:2011aa}. Scale factors are applied as event weights to the simulated samples to correct for the differences in the efficiencies measured with the data and the simulation. The scale factors for trigger, reconstruction, identification, and isolation efficiencies depend on  $\pt$ and $\abs{\eta}$. For the dielectron channel the trigger efficiency scale factors range from 0.92 to 1.03  with an uncertainty in the range 0.1 to 1.9\%. The reconstruction efficiency scale factors vary from 0.98 to 1.01 with uncertainties of 0.1 to 1.2\% respectively, while the combined identification and isolation efficiency scale factors range from 0.91 to 1.02 with uncertainties of 0.1 to 5.7\%. For the dimuon channel the scale factor for the trigger efficiency varies from 0.97 to 1.01 with a typical uncertainty of 0.2\%,  and the combined scale factor for the reconstruction, identification, and isolation efficiencies ranges from 0.92 to 1.03 with an uncertainty of about 0.5\%. Energy and momentum scale corrections are applied to the electrons and muons, respectively, in both experimental data and simulated events~\cite{Bodek:2012id,Chatrchyan:2013mxa}.

Thirty-four bins in $\phi^*$ are defined~\cite{Aad:2015auj} with widths that increase  with $\phi^*$; the bulk of the distribution falls in the range $\phi^* < 1$.    When measuring the double-differential cross section, six bins in $\abs{y}$ of constant  width $\Delta\abs{y} = 0.4$ covering the range $\abs{y} < 2.4$ are used.

 \section{Background estimation}
 \label{sec:bkgnd}

The background contributions to the selected samples amount only to about~0.6\% and~0.5\%
in the dielectron and dimuon channels, respectively. The components of this background consist
of the inclusive production of \ttbar, $ \cPZ\to\TT$, $\PW\PW$, $\PW\Z$, $\Z\Z$,  single top quarks, and, to a lesser extent,  \PW+jets and
QCD multijets. The latter two processes contribute when at least one jet is misidentified as a lepton or when a lepton produced within a jet passes the isolation requirement. Their contribution in the dimuon channel is negligible. In the dielectron channel the background arising from \PW+jets and QCD multijet processes is estimated by fitting the invariant mass distribution in each final bin.  The fit is performed using an analytical shape for the \PW+jets and QCD multijet backgrounds and a simulation-derived shape for the other backgrounds and the signal events that have wrongly reconstructed same-sign dielectrons.  Since the processes which generate dielectron pairs in QCD multijets and \PW+jets are expected to be charge-symmetric, the analytical fit result from the same-sign distribution is used to predict the background in the total sample. This background constitutes approximately 6\% of the total background in the dielectron channel.
All other backgrounds are estimated using simulated event samples.
As indicated in Eqs.~(\ref{eq1}) and~(\ref{eq2}), the estimated total background is subtracted bin-by-bin before unfolding the spectra.

\begin{figure}[htbp]
  \centering
    \includegraphics[width=0.44\textwidth]{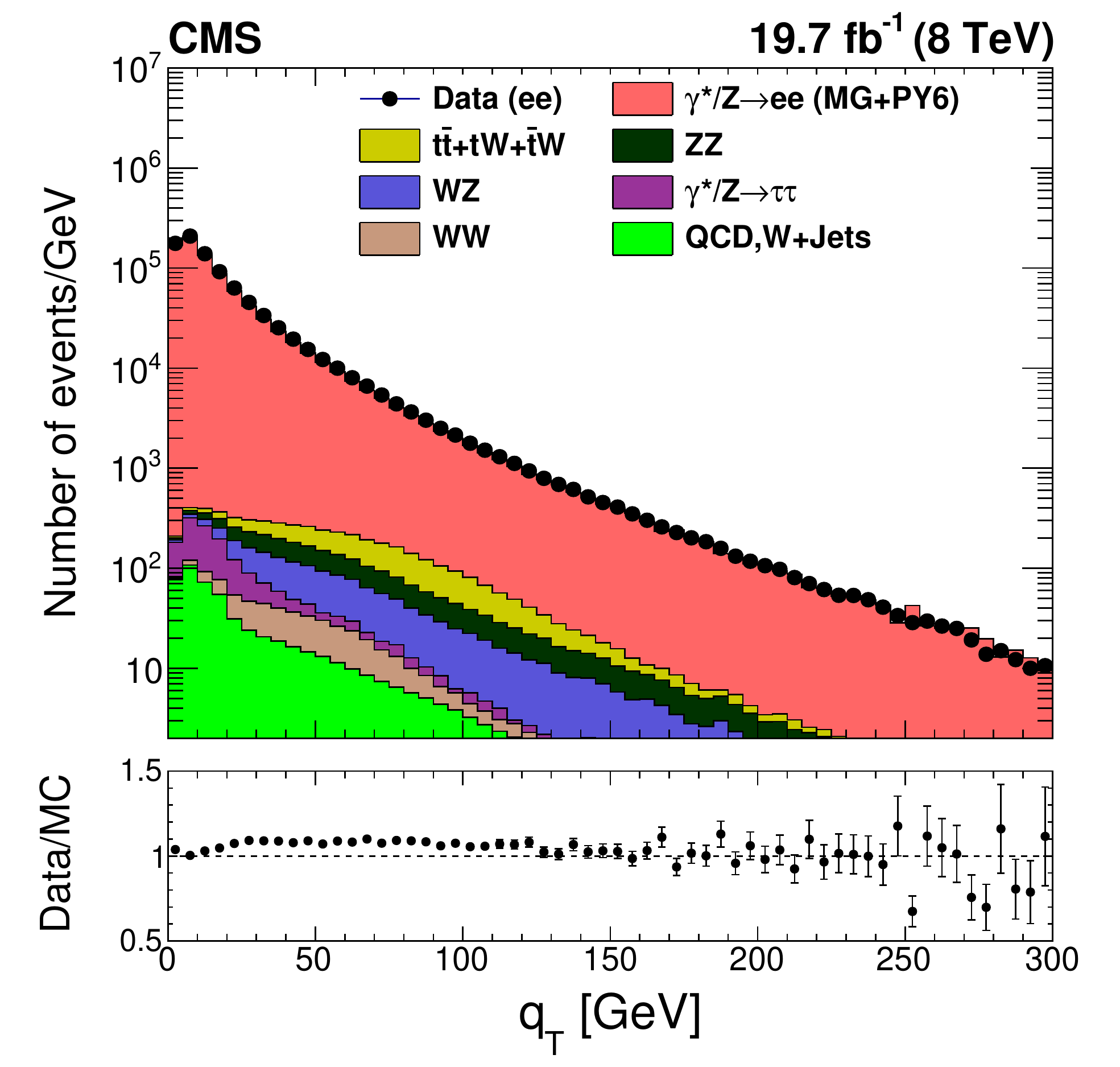}
    \includegraphics[width=0.44\textwidth]{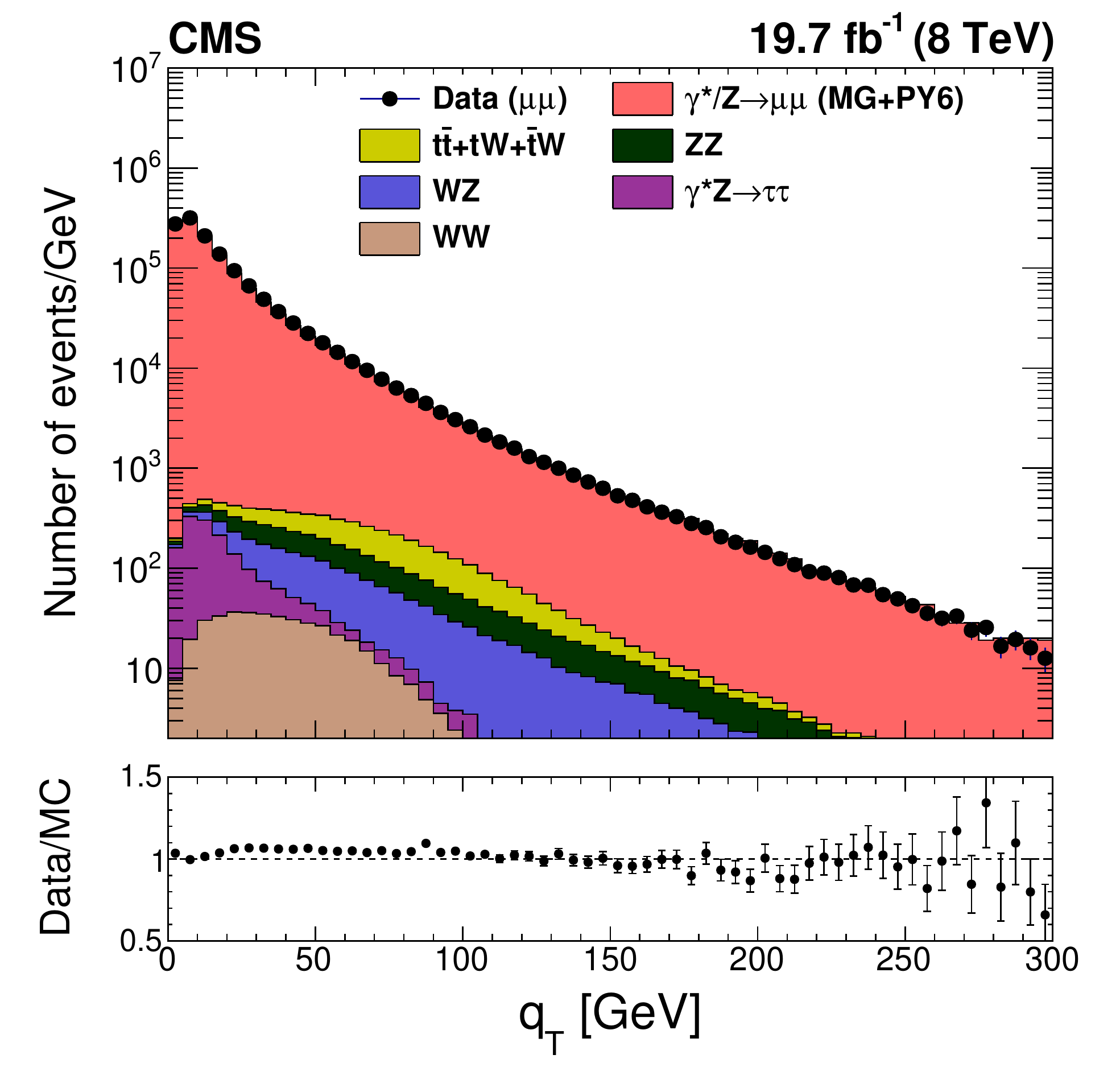}
    \includegraphics[width=0.44\textwidth]{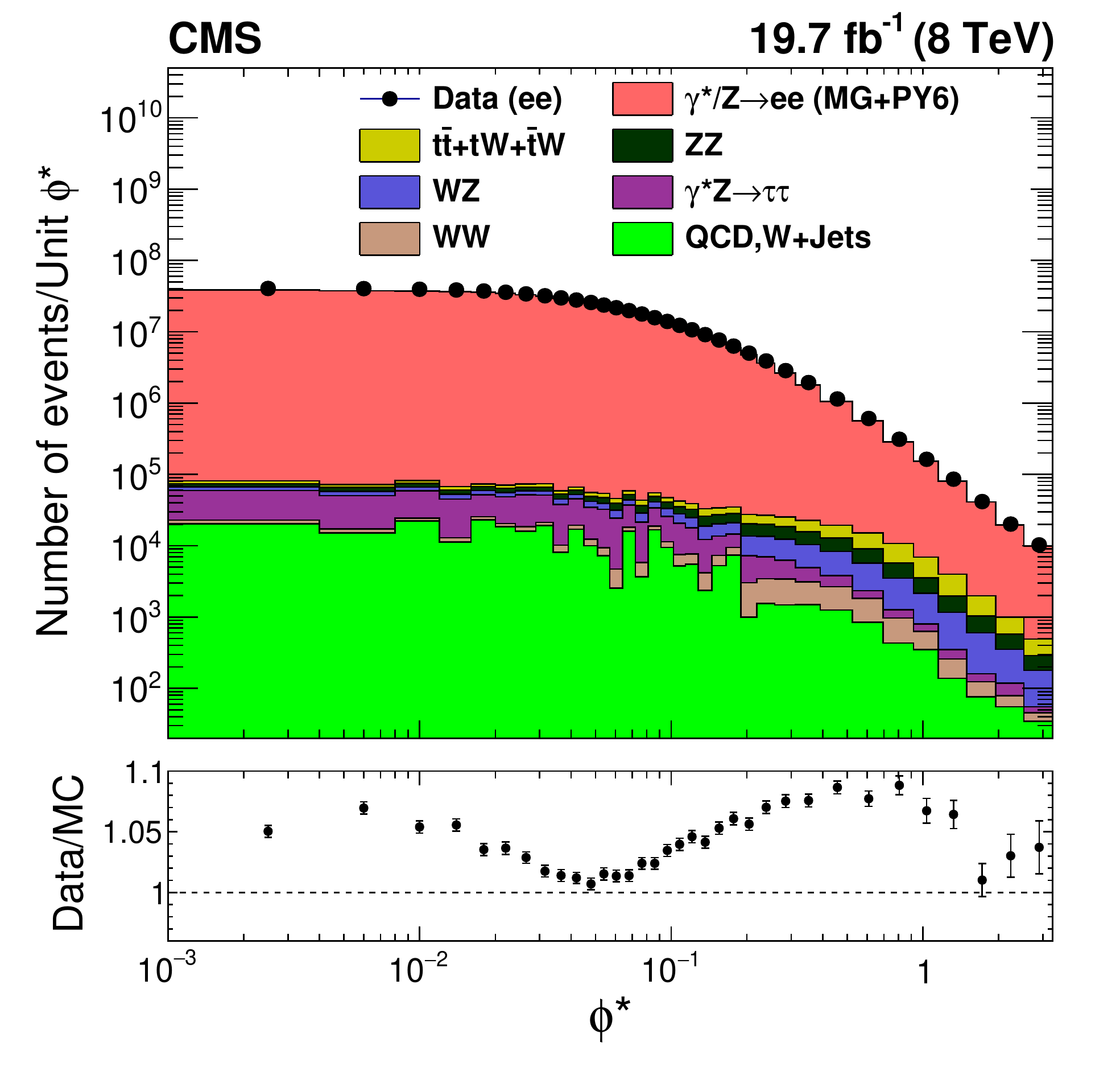}
    \includegraphics[width=0.44\textwidth]{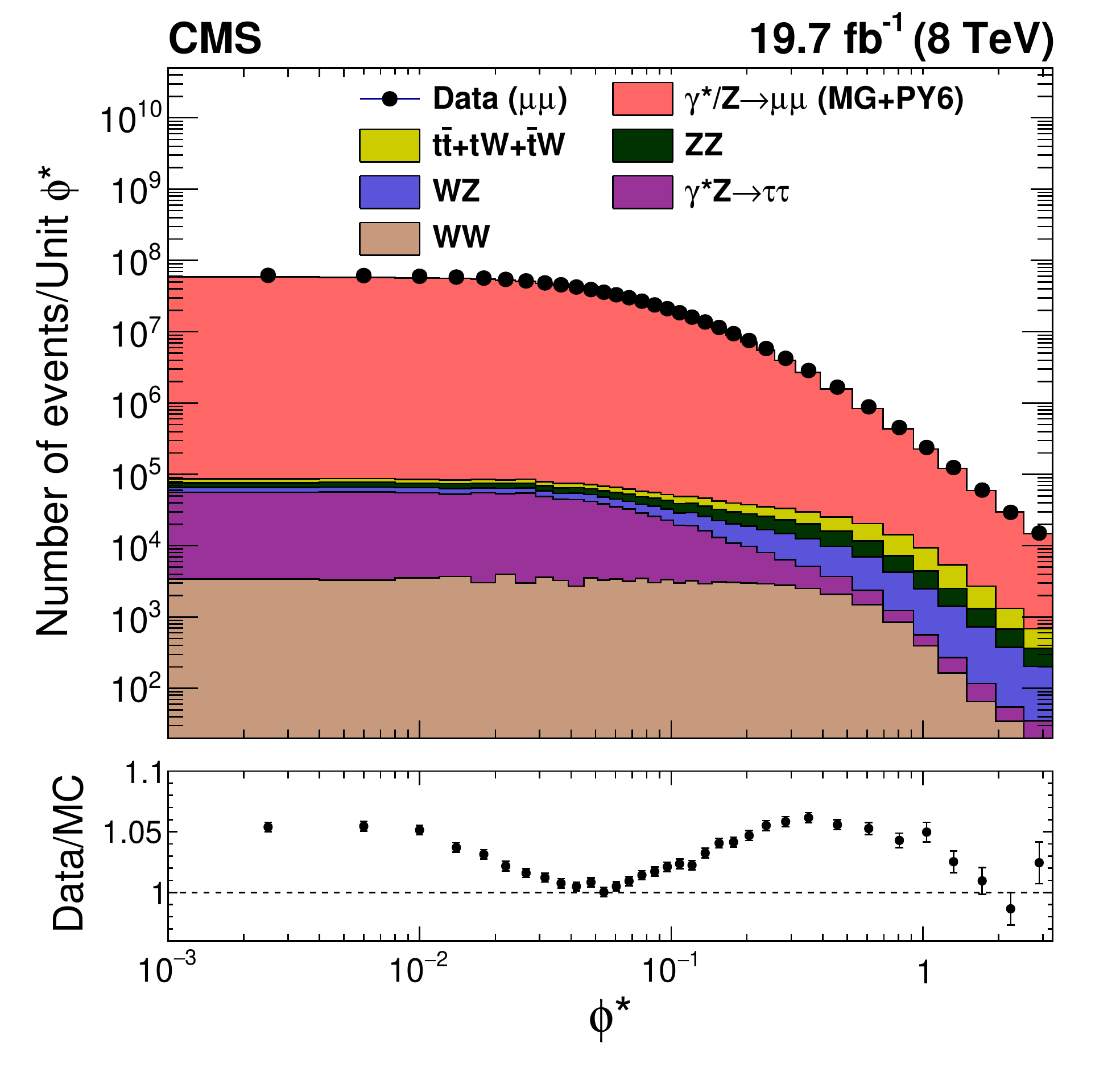}
    \includegraphics[width=0.44\textwidth]{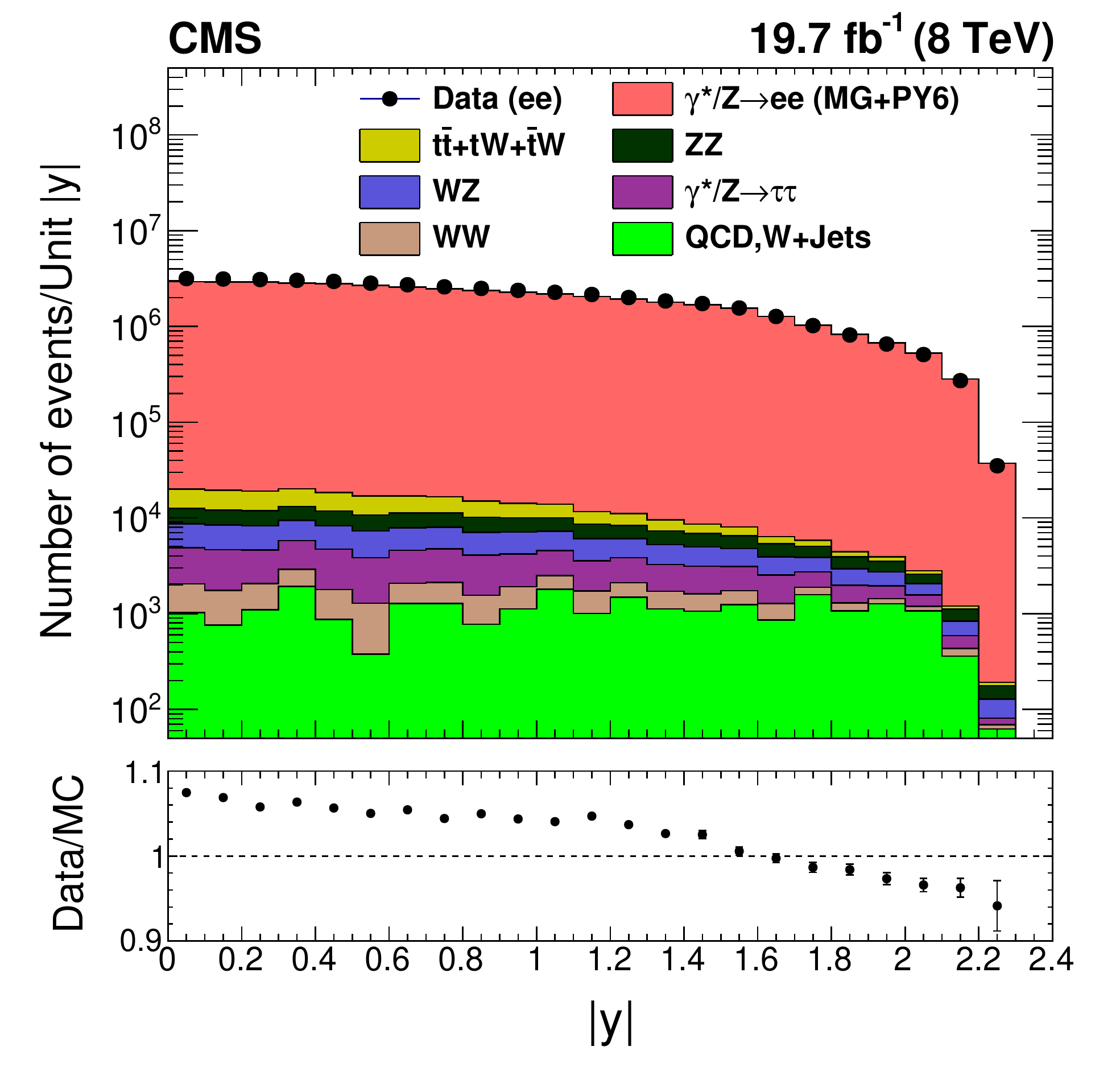}
    \includegraphics[width=0.44\textwidth]{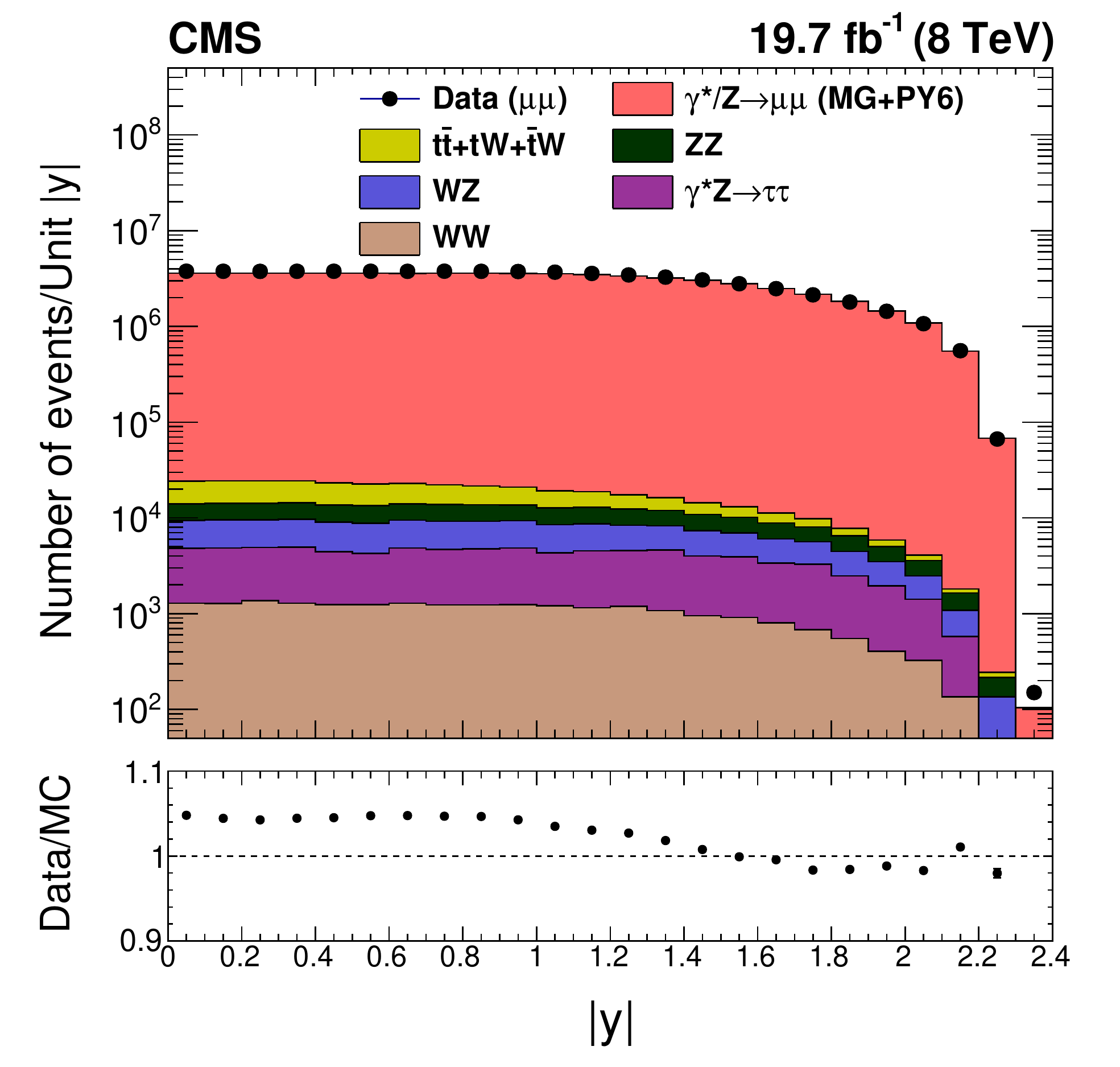}
    \caption{Distributions of dilepton transverse momentum $\qT$ (upper), $\phi^*$~(middle),  and rapidity~$\abs{y}$ (lower) in the dielectron~(left) and dimuon~(right) channels.
     The points represent the data and the shaded histograms represent the expectations which
     are based on simulation, except for the contributions from QCD multijet and \PW+jets events in the dielectron channel,
     which are obtained from control samples in data. Here ``MG+PY6'' refers to  a sample produced with \MADGRAPH interfaced with \PYTHIASIX~(Z2* tune).
     The error bars indicate the  statistical uncertainties for data and for simulation only.
     No unfolding procedure has been applied to these distributions.}
    \label{fig:zPtP}

\end{figure}

Figure~\ref{fig:zPtP} presents the observed and the expected dielectron and dimuon kinematic
distributions.  Scale factors have been applied to remove any differences in efficiency between
data and simulation as discussed earlier; weights have been applied to match the distribution of pileup vertices in data.
The error bars represent the statistical uncertainties for the data and the simulations.
The top row displays the $\qT$ distribution followed by the $\phi^*$ and $\abs{y}$ distributions.
The data and the expectations in all distributions agree within 10\%.

 \section{Unfolding}
 \label{sec:unfolding}

To compare with the predictions from event generators, the distributions of the observables need to be corrected back to the stable particle level for event selection efficiencies and for detector resolution effects. The measurement uncertainties for $\phi^*$ and $\abs{y}$ are small, but not zero.
In order to remove the impact of events migrating among bins, the background-subtracted distributions are unfolded. For the double-differential distribution, the migration of events from one $\phi^*$ bin to another  is at the level of 10\ (3) \% for the dielectron (dimuon) channel, while for the $\abs{y}$ distribution the corresponding values are  smaller, typically less than 2 (1)\%, because the $\abs{y}$~bins are large compared to the resolution.
In addition to the effects of measurement uncertainties,
the impact of QED-FSR is included in the unfolding.
The observed distributions are unfolded to pre-FSR or ``Born-level''
distributions using the
d'Agostini method~\cite{1995NIMPA.362..487D}
as implemented in the \textsc{RooUnfold} package~\cite{Adye:2011gm}.
Four iterations have been performed for the unfolding of the distributions. A response matrix  correlates the values of the observable with
and without the detector effects.
The model for the detector resolution is derived from
a simulated signal sample generated with \MADGRAPH interfaced with \PYTHIASIX.

 \section{Systematic uncertainties}
 \label{sec:systematics}

\begin{figure}[htbp]
  \centering
    \includegraphics[width=0.48\textwidth]{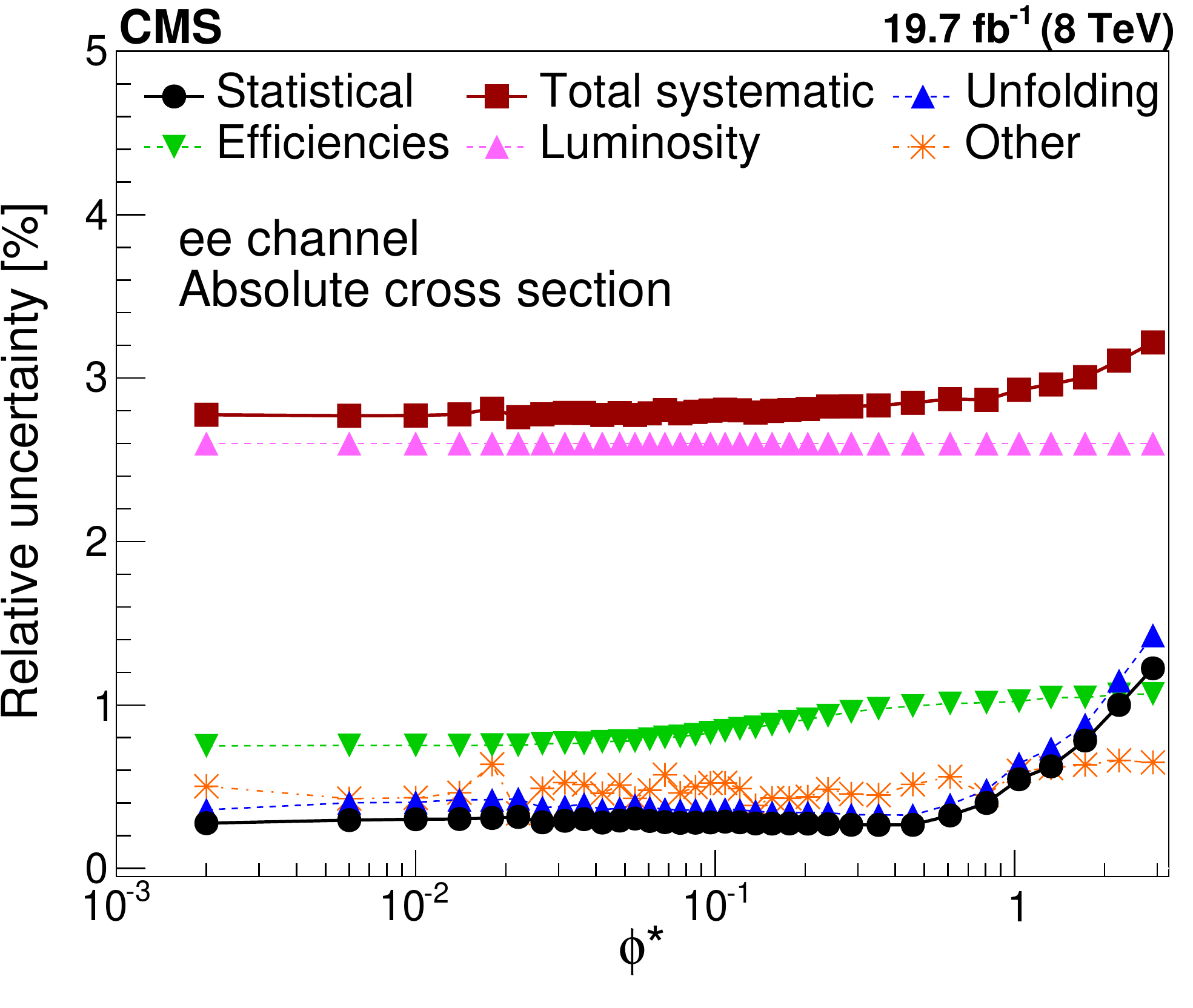}
    \includegraphics[width=0.48\textwidth]{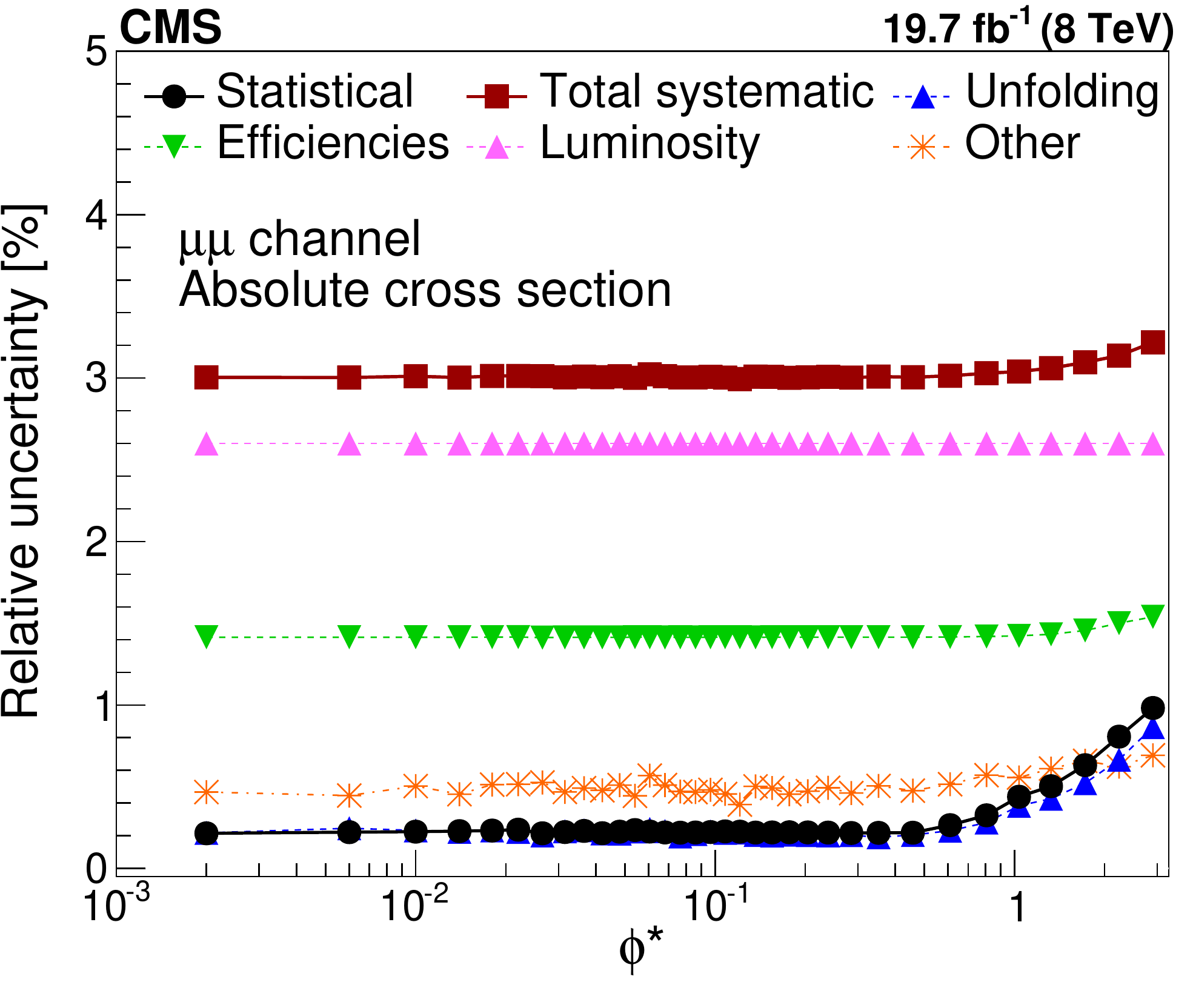}
    \includegraphics[width=0.48\textwidth]{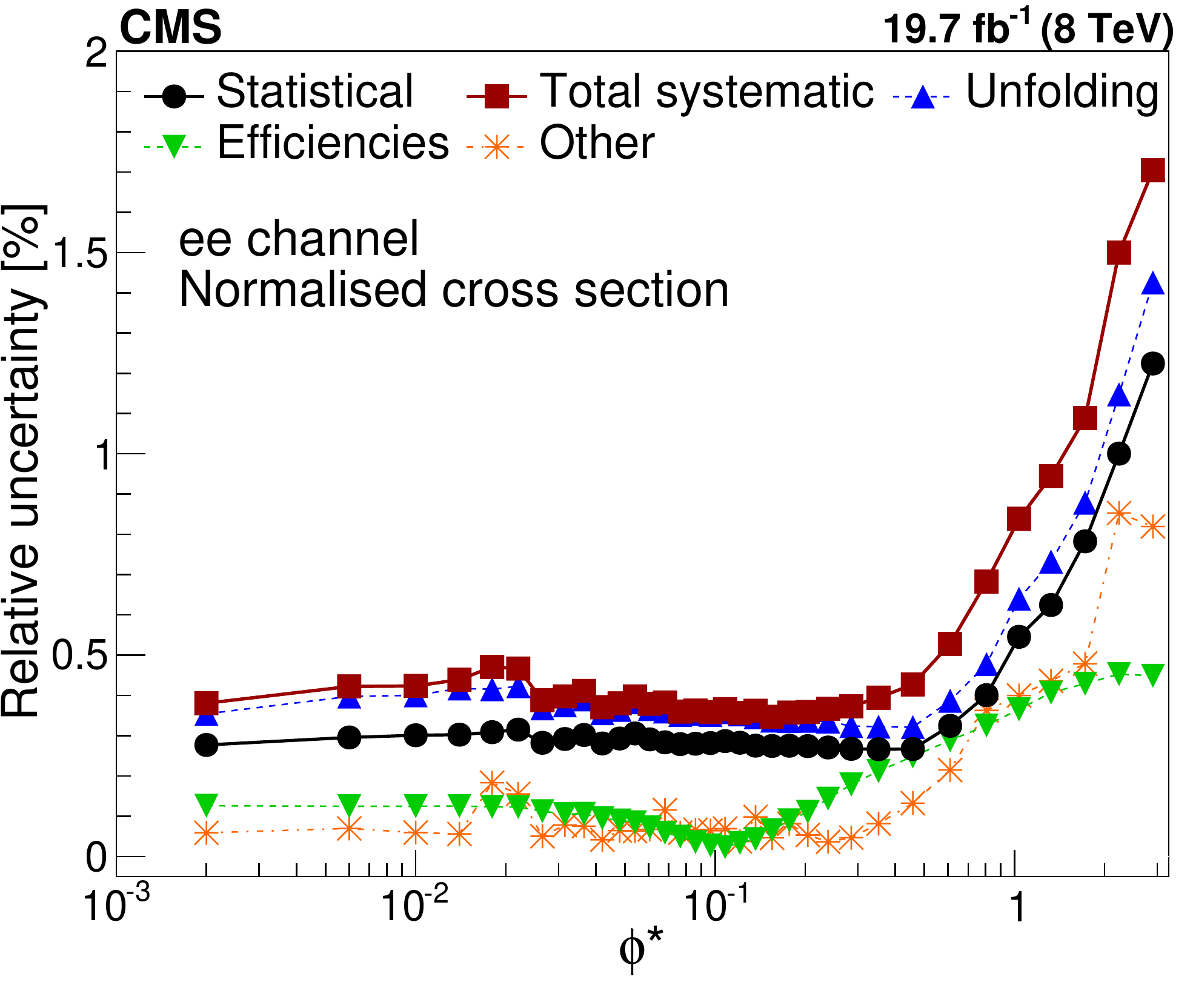}
    \includegraphics[width=0.48\textwidth]{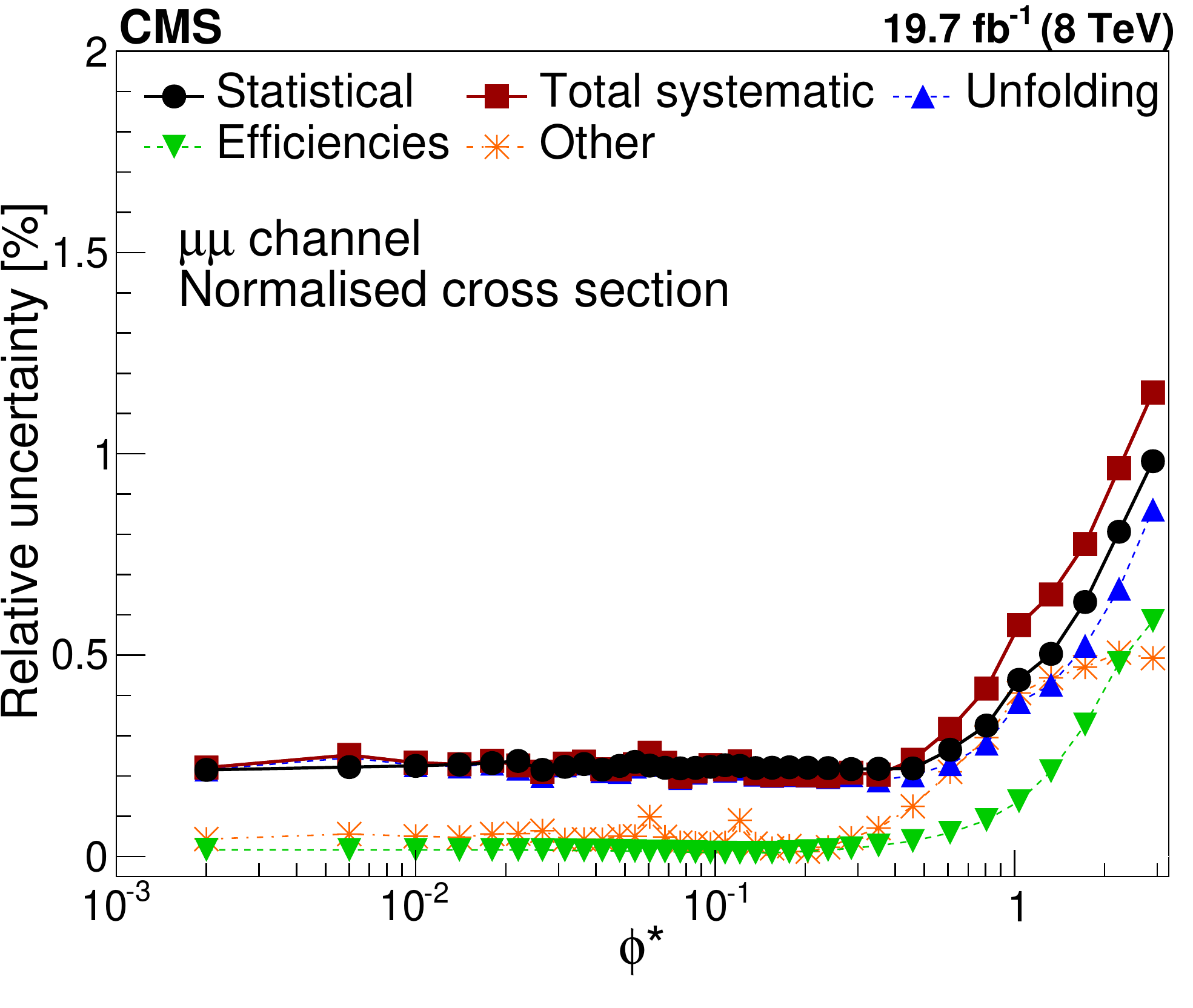}
    \caption{The variation of statistical and systematic uncertainties
      with~$\phi^*$.  The upper row shows the relative uncertainty
      for the absolute cross section while the lower one shows the relative
      uncertainty for the normalised cross section.
      The left plots pertain to the dielectron channel and the right plots
      pertain to the dimuon channel.
      The uncertainties from the background, pileup, the electron energy scale or the
      muon $\pt$ resolution, and from  QED-FSR modelling are combined under the label ``Other''.}
    \label{fig:syst}

\end{figure}

The total systematic uncertainty includes uncertainties in the integrated luminosity, unfolding,
lepton efficiencies (trigger, identification and isolation), pileup, background estimation,
electron energy scale,
muon momentum scale and resolution,
and modelling of QED-FSR. The impact of these sources of systematic uncertainty varies with~$\phi^*$,
as shown in Fig.~\ref{fig:syst}, and is different for the measurement of
absolute and normalised cross sections. As expected, the systematic uncertainties
for the normalised cross sections
are substantially smaller than those for the absolute cross section.

The largest source of uncertainty comes from the measurement of the integrated luminosity
and amounts to~2.6\%~\cite{CMS-PAS-LUM-13-001}.  It is uniform across all~$\phi^*$
and $\abs{y}$ bins and is relevant only for the absolute cross section measurements.

The unfolding uncertainty originates from the finite size of the simulated signal sample used for the response matrix and hence the variation of this uncertainty with
$\phi^*$ and $\abs{y}$ closely parallels the statistical uncertainty.
The model dependence is studied by reweighting the simulated events used for the unfolding
to match either the $y$ or $\Mll$ distribution in data
or to change the $\qT$ distribution. The effect of this reweighting on the unfolded data
is less than 0.05\% for most of the $\phi^*$ range and reaches about 0.5\% for the highest bin of the $\abs{y}$  distribution. The systematic uncertainty due to the model dependence of the unfolding procedure is of comparable magnitude and both are negligible.
Systematic uncertainties for lepton efficiencies include the uncertainties in the scale
factors used to correct the identification, isolation, and trigger efficiency values
from the simulation.

The uncertainty in the background estimates from the  simulated samples is assessed by varying the cross sections of the  contributing processes by the amount as measured by the CMS Collaboration.  The \ttbar background is varied by 10\%~\cite{Khachatryan:2016mqs}
while~$\PW\Z$ and~$\Z\Z$ contributions are varied simultaneously by
20\%~\cite{Khachatryan:2016poo,CMS:2014xja}. In the dielectron channel the
contribution due to  QCD multijets and \PW+jets processes
is assigned a conservative uncertainty of~100\% based on variations observed when the binning is changed.
Uncertainties in the other background processes lead to negligible effects on the measured cross sections, being less than a tenth of the effect of the major backgrounds.

The electron energy scale, known to a precision of 0.1--0.2\%, affects
all of the $\phi^*$ bins almost uniformly
at the level of~0.15\% for the absolute cross section measurement.
The impact on the normalised cross sections is smaller, at the level of~0.06\%.
 The muon momentum scale
is corrected for the misalignments in the detector systems and the uncertainty in the knowledge of the magnetic field.
The corresponding cross section uncertainties are below 0.1\% level.

To account for the uncertainty in  QED-FSR, the simulation is weighted to reflect the difference  between a soft-collinear approach and the exact $\mathcal{O}(\alpha)$ result as obtained in PHOTOS~\cite{Nanava:2003cg}. This uncertainty is less  than~0.08\% in the entire phase space considered.

To estimate the uncertainty in our measurement due to that in pileup multiplicity, the number of interactions per bunch crossing in the simulation is varied by~$\pm$5\%. This includes the effects due to the modelling of minimum bias events in simulation, uncertainty in the measurement of the inelastic cross section and the number of interactions per bunch crossing as measured in data.

The uncertainty in the cross sections due to variations of the structure functions in the used PDF sets is negligible.

\begin{figure}[tp]
  \centering
    \includegraphics[width=0.48\textwidth]{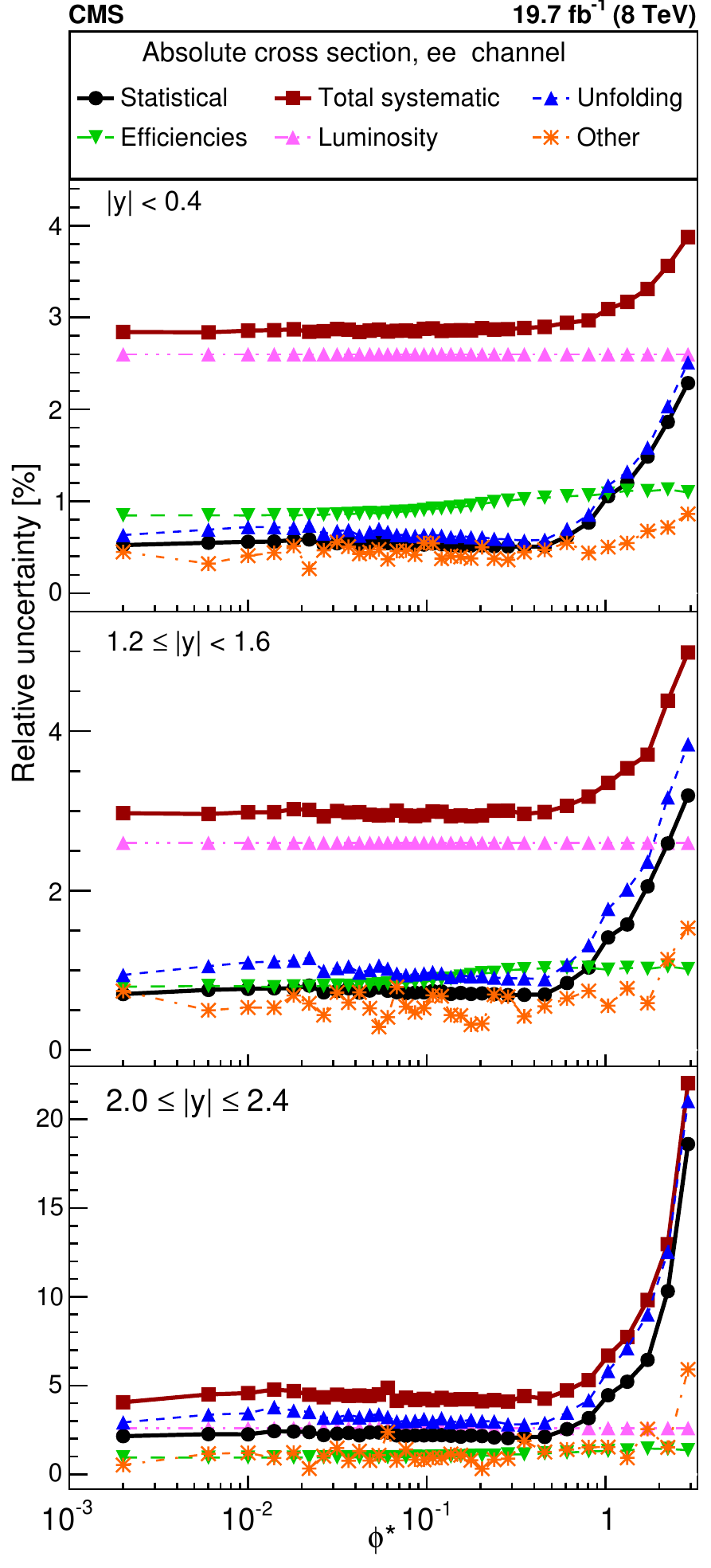}
    \includegraphics[width=0.48\textwidth]{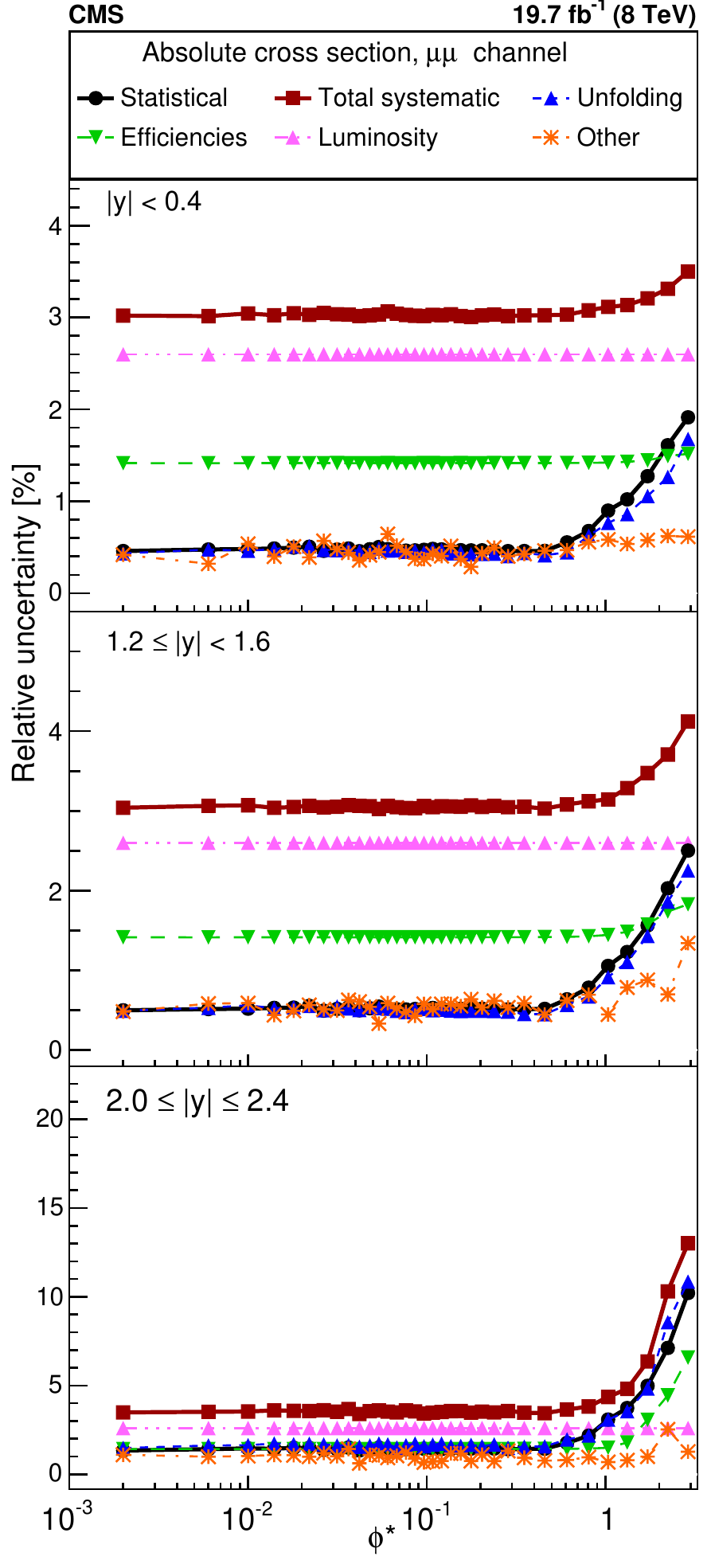}
    \caption{The variation of statistical and systematic uncertainties, in representative $\abs{y}$ bins, for the ${\rd}^2 \sigma / {\rd} \phi^* {\rd}\abs{y}$ measurements,  in the dielectron (left) and dimuon (right) channels.  The main components are shown individually while uncertainties from the background, pileup, the electron energy scale or the muon $\pt$ resolution, and from  QED-FSR are combined under the label ``Other''.}
    \label{fig:UncAbs2d}

\end{figure}

\begin{figure}[htbp]
  \centering
    \includegraphics[width=0.48\textwidth]{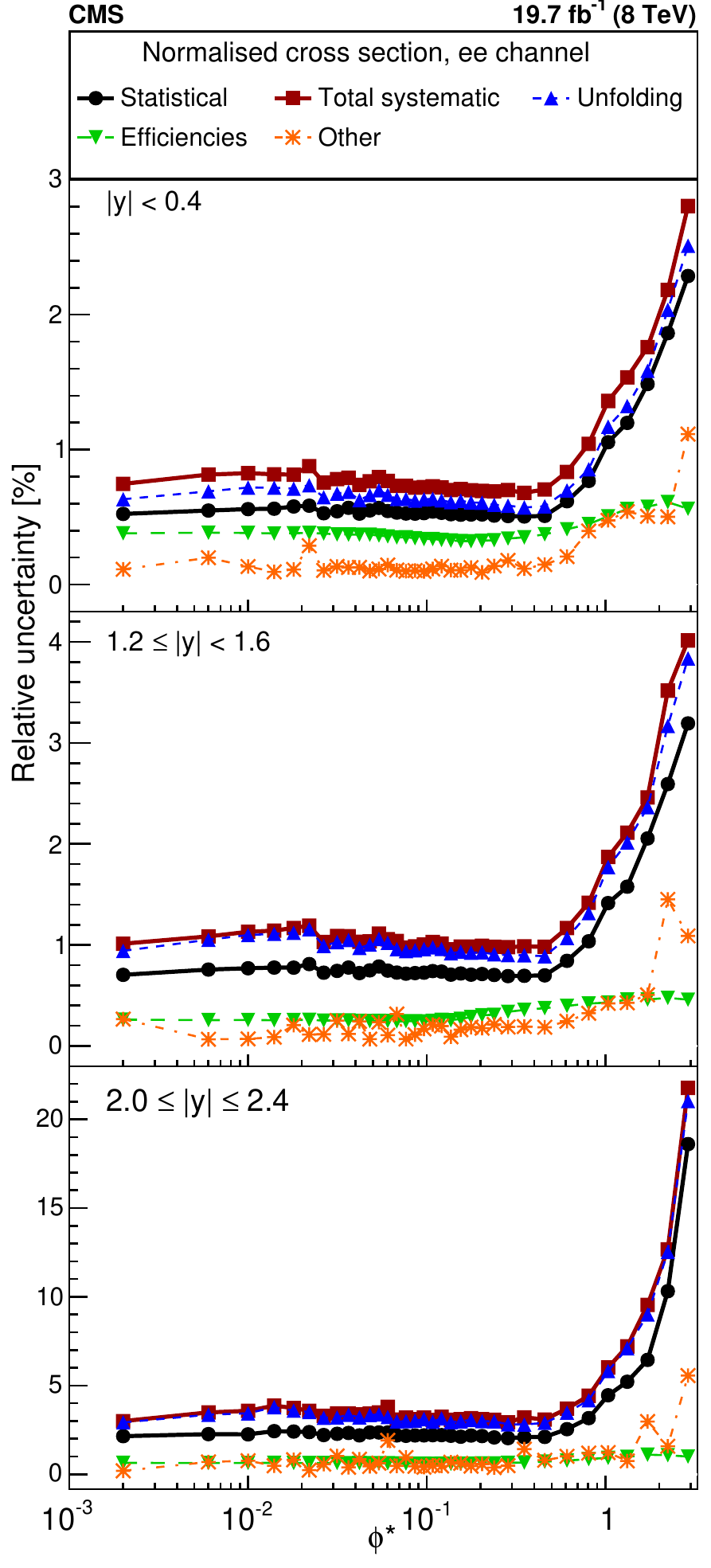}
    \includegraphics[width=0.48\textwidth]{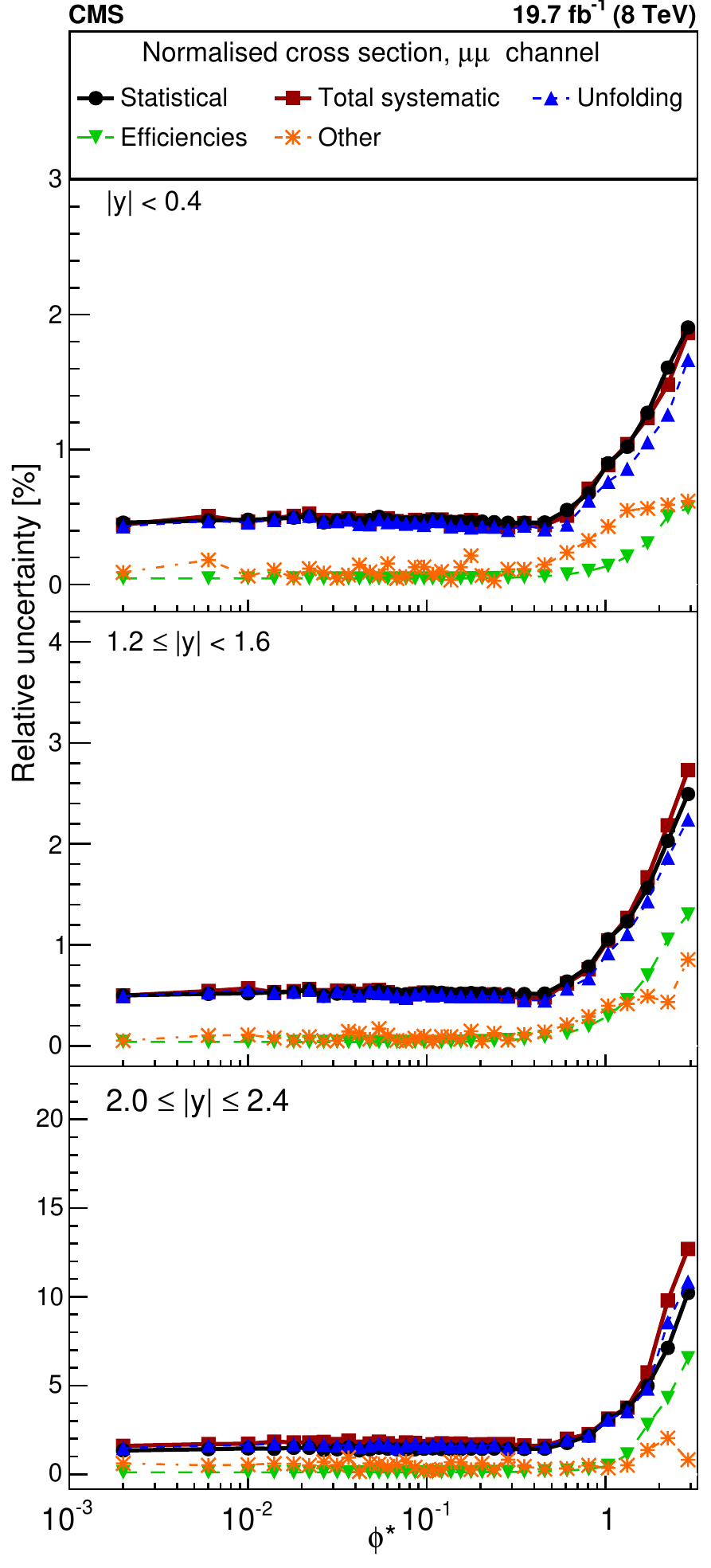}
    \caption{The variation of statistical and systematic uncertainties, for the normalised double-differential cross section measurements, in representative $\abs{y}$ bins, in the dielectron (left) and dimuon (right) channel. The main components are shown individually while uncertainties from the background, pileup, the electron energy scale or the muon $\pt$ resolution, and from QED-FSR are combined under the label ``Other''.}
    \label{fig:UncNorm2d}

\end{figure}

Summaries of the uncertainties for the absolute and normalised double-differential
cross section measurements and their variations with~$\phi^*$
in representative $\abs{y}$ bins
are displayed in Figs.~\ref{fig:UncAbs2d} and \ref{fig:UncNorm2d}, respectively.
For the double-differential cross section, the statistical uncertainty from the data and
the MC unfolding statistical uncertainty
are larger than in the single-differential cross section measurement.
The statistical uncertainty starts to dominate the total uncertainty in the high~$\phi^*$
and high-$\abs{y}$ regions. Furthermore, the relative contribution of
the background processes in the fiducial region, and therefore
the background uncertainty, increases with rapidity.
This is especially true for the QCD multijet and \PW+jets backgrounds in the dielectron
channel, leading to an uncertainty of approximately~5\% in the highest ranges of  $\phi^*$
and~$\abs{y}$ covered, which nonetheless remains smaller than the statistical uncertainty.

 \section{Theoretical predictions}

\label{sec:predictions}

The measured differential cross sections are compared with five theoretical predictions.
Apart from the LO predictions of \MADGRAPH described in Section~\ref{sec:datamc_description}, the following are also considered:  (i) \POWHEG~\cite{Nason:2004rx,Alioli:2010xd,Alioli:2010qp,Frixione:2007vw}  with the CT10NLO PDFs~\cite{Gao:2013xoa} interfaced with \PYTHIASIX and the Z2* tune; (ii) \POWHEG  with the CT10NLO PDF, but interfaced with \PYTHIAEIGHT (v8.2)~\cite{Sjostrand:2014zea}  and the CUETP8M1 tune~\cite{Khachatryan:2015pea} using NNPDF2.3 LO PDF~\cite{Ball:2010de, Ball:2011mu}; (iii) \RESBOS~\cite{Ladinsky:1993zn, Balazs:1997xd, Landry:2002ix} with CT10NLO PDF, and (iv) \AMCATNLO\ (henceforth referred to as \aMCATNLO{})~\cite{Alwall:2014hca} with the NNPDF3.0 NLO PDF and \PYTHIAEIGHT for the parton shower and FxFx merging scheme~\cite{Frederix:2012ps}. The generators \POWHEG and \aMCATNLO{} are both accurate at NLO, while the order for \RESBOS is resummed NNLL/NLO QCD. Since \RESBOS uses the resummation method of $\pt$ to account for contributions from soft-gluon radiations in the initial state it differs from fixed-order perturbative calculations and MC showering methods. \RESBOS predictions have been obtained with $CP$ version using general purpose grids.

The \MADGRAPH predictions are normalised to the \FEWZ cross section for $\Mll > 50\GeV$~\cite{Li:2012wna}.
The uncertainties in the total theoretical cross section calculated with \FEWZ
include those due to $\alpS$, neglected higher-order QCD terms beyond NNLO,
the choice of heavy-quark masses (bottom and charm), and PDFs, amounting to a total of~3.3\%.
The theoretical uncertainties for \POWHEG, \RESBOS, and \aMCATNLO include statistical, PDF, and scale uncertainties.
The PDF uncertainty is calculated using the recommendations of Ref.~\cite{Alekhin:2011sk,Botje:2011sn}, and
the scale uncertainties are evaluated by varying the renormalisation and the factorisation scales
independently by factors of 2 and 1/2 and taking the largest variations as the uncertainty.

 \section{Results}
 \label{sec:results}

The measurements in the dielectron and dimuon channels are consistent within
the uncorrelated statistical and systematic uncertainties, and hence they are combined.   The best linear unbiased estimator (BLUE) method~\cite{blue1,Valassi:2003mu}, as implemented in Ref.~\cite{BLUE2}  is used. The resulting output is unbiased and has minimal variance.  The correlations among bins in one channel as well as  between the two channels, including those in the unfolding, are taken into account. The correlation between channels originates from the systematic uncertainties due to background estimates, pileup, QED-FSR, and the integrated luminosity.
The correlations within one channel also include uncertainties from the lepton
efficiencies. The uncertainty in the integrated luminosity  is fully correlated
across all bins and both final states. It is evaluated for the final result after  combining channels with the BLUE method.

\begin{figure}[htbp]
\centering
\includegraphics[width=0.6\textwidth]{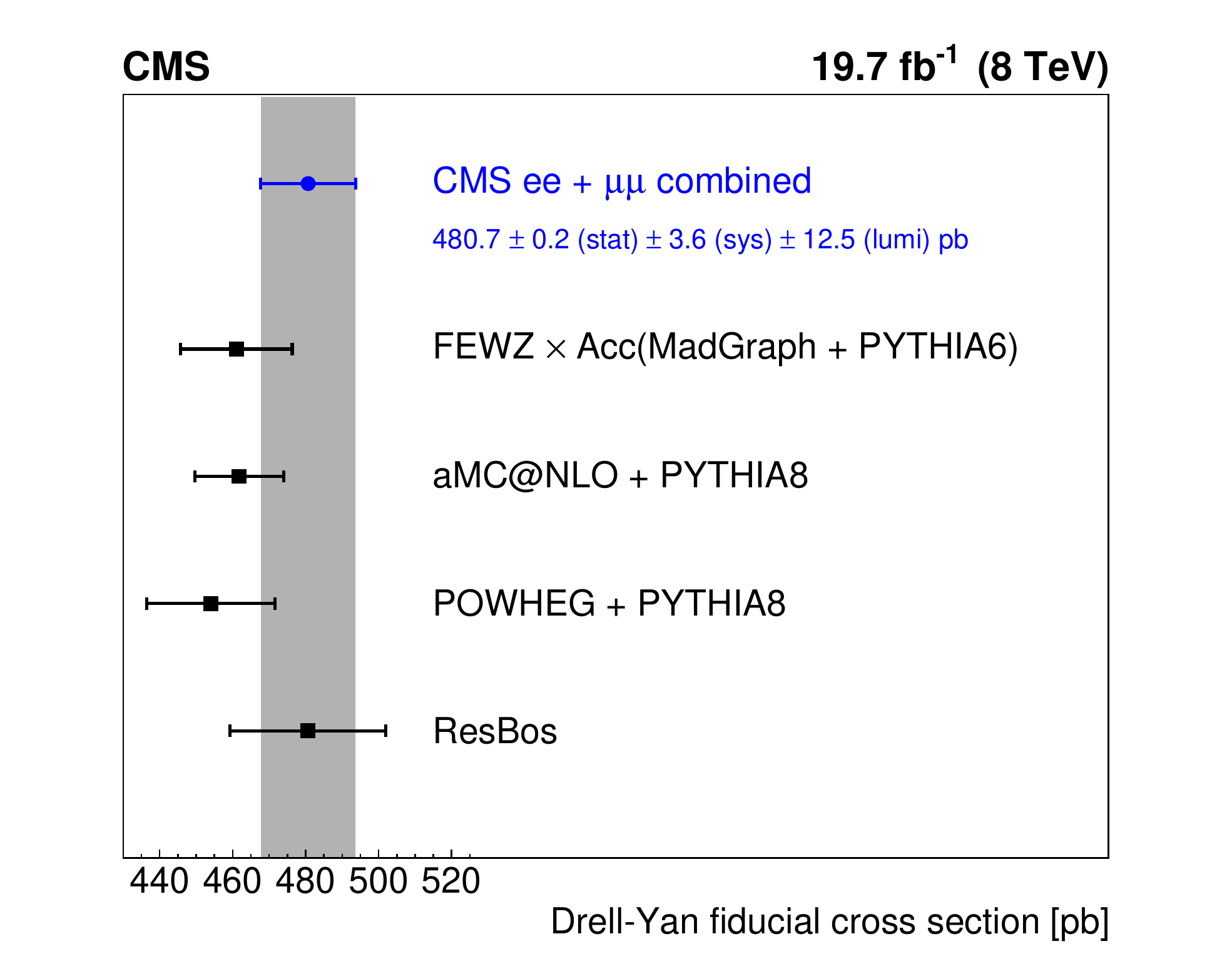}
\caption{\label{fig:sigmafid}
Comparison of theoretical values for the fiducial cross section
with the measured value.
The grey error bar represents the total experimental uncertainty for the
measured value.  The error bars for the theoretical values include
the uncertainties due to statistical precision, the PDFs,
and the scale choice.
The fiducial cross section for \FEWZ  is obtained by multiplying the total cross section with  the acceptance determined from the simulated \MADGRAPH{}+\PYTHIASIX sample; the uncertainty in the prediction corresponds to that in the \FEWZ calculation.
}
\end{figure}

The fiducial cross section, as defined in Section~\ref{sec:method},
is obtained by integrating the absolute
differential cross section $\ddinline{\sigma}{\phi^*}$.
After combining dielectron and dimuon channels,
the measured value for a single lepton flavour is
\begin{equation}
\label{eq:sigmafid}
 \sigma (\Pp\Pp \to \cPZ/\gamma^* \to \ell^+\ell^-) = 480.7 \pm 0.2\stat \pm 3.6\syst\pm 12.5\lum\unit{pb},
\end{equation}
where the statistical, systematic, and integrated luminosity uncertainties are indicated separately.
As shown in Fig.~\ref{fig:sigmafid}, this measurement is in agreement with the theoretical predictions which have a typical uncertainty of 3\%.

\begin{figure}[htbp]
\centering
\includegraphics[width=0.48\textwidth]{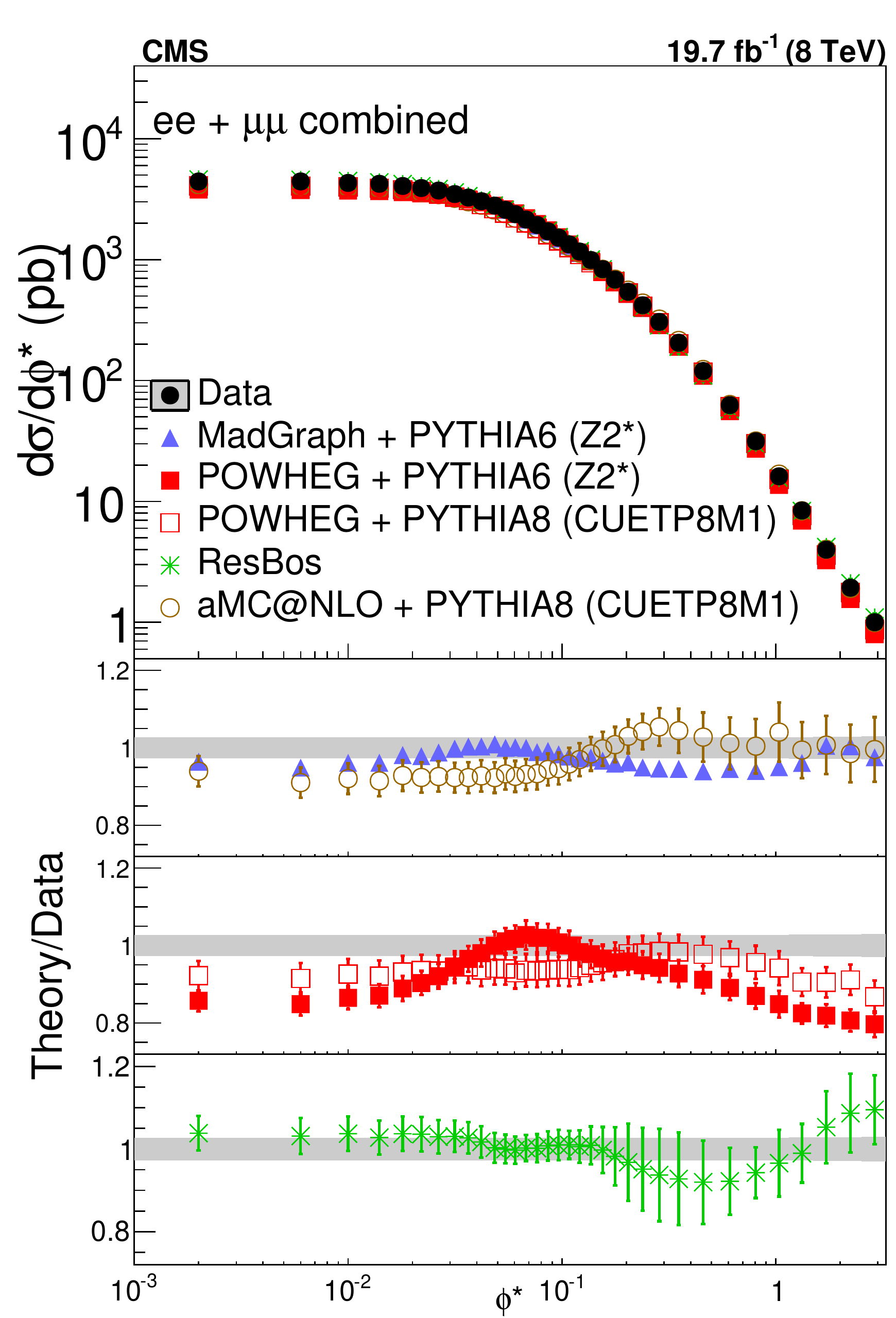}
\includegraphics[width=0.48\textwidth]{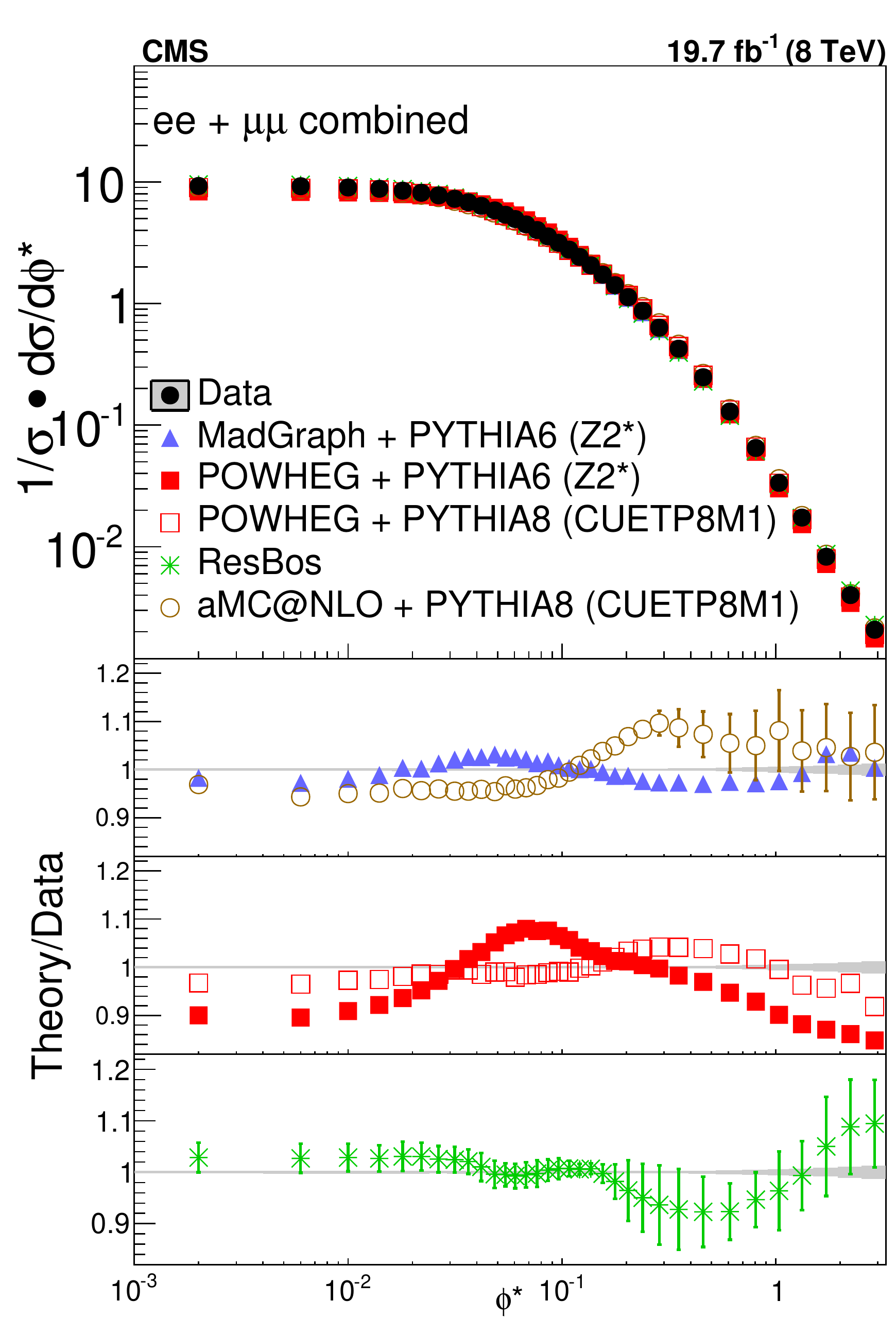}
\caption{The measured absolute (left) and the normalised (right) cross sections after the combination of dielectron and dimuon channels. The measurement is
compared with the predictions from \RESBOS, \MADGRAPH and \POWHEG interfaced with \PYTHIASIX~(Z2* tune), and \aMCATNLO{} and \POWHEG interfaced with \PYTHIAEIGHT~(CUETP8M1 tune). In the lower panels, the horizontal  bands  correspond to the experimental uncertainty, while the error bars correspond to the statistical,  PDF, and scale uncertainties in the theoretical predictions from \RESBOS,  \POWHEG and \aMCATNLO and only the statistical uncertainty for \MADGRAPH.}
\label{fig:All}
\end{figure}

{\tolerance=2400
The combined absolute and normalised single-differential cross sections, $\ddinline{\sigma}{\phi^*}$ and
$(1 / \sigma)\,\ddinline{\sigma}{\phi^*}$  are presented in Fig.~\ref{fig:All}.
The lower panels indicate the conformity of theory with data.
None of the predictions matches the measurements perfectly
for the entire range of~$\phi^*$ covered in this analysis.
For the normalised cross section, \MADGRAPH{}+\PYTHIASIX provides
the best description with a disagreement of at most 5\% over the entire $\phi^*$ range.
\RESBOS, \aMCATNLO{}+\PYTHIAEIGHT and \POWHEG{}+\PYTHIAEIGHT predictions are similarly successful at describing
the data at low $\phi^*$ but they disagree with the measurements by as much as 10\% for $\phi^* > 0.1$.
\POWHEG{}+\PYTHIASIX provides the least accurate prediction, with a disagreement
up to 11\,(15)\% for $\phi^*$ less (greater) than value 0.1.
Better models of the hard-scattering process, such as provided by \MADGRAPH{}+\PYTHIASIX,
lead to an improved agreement with the data.  At the same time,
the importance of the underlying event model and hadronisation tune for correctly reproducing
the $\phi^*$~distribution is evident from the  significant difference (up to 11\%) in
predicted distributions for a given sample of \POWHEG events hadronised with
\PYTHIASIX and with \PYTHIAEIGHT separately.
\par}

\begin{figure}[htbp]
\centering
\includegraphics[width=0.48\textwidth]{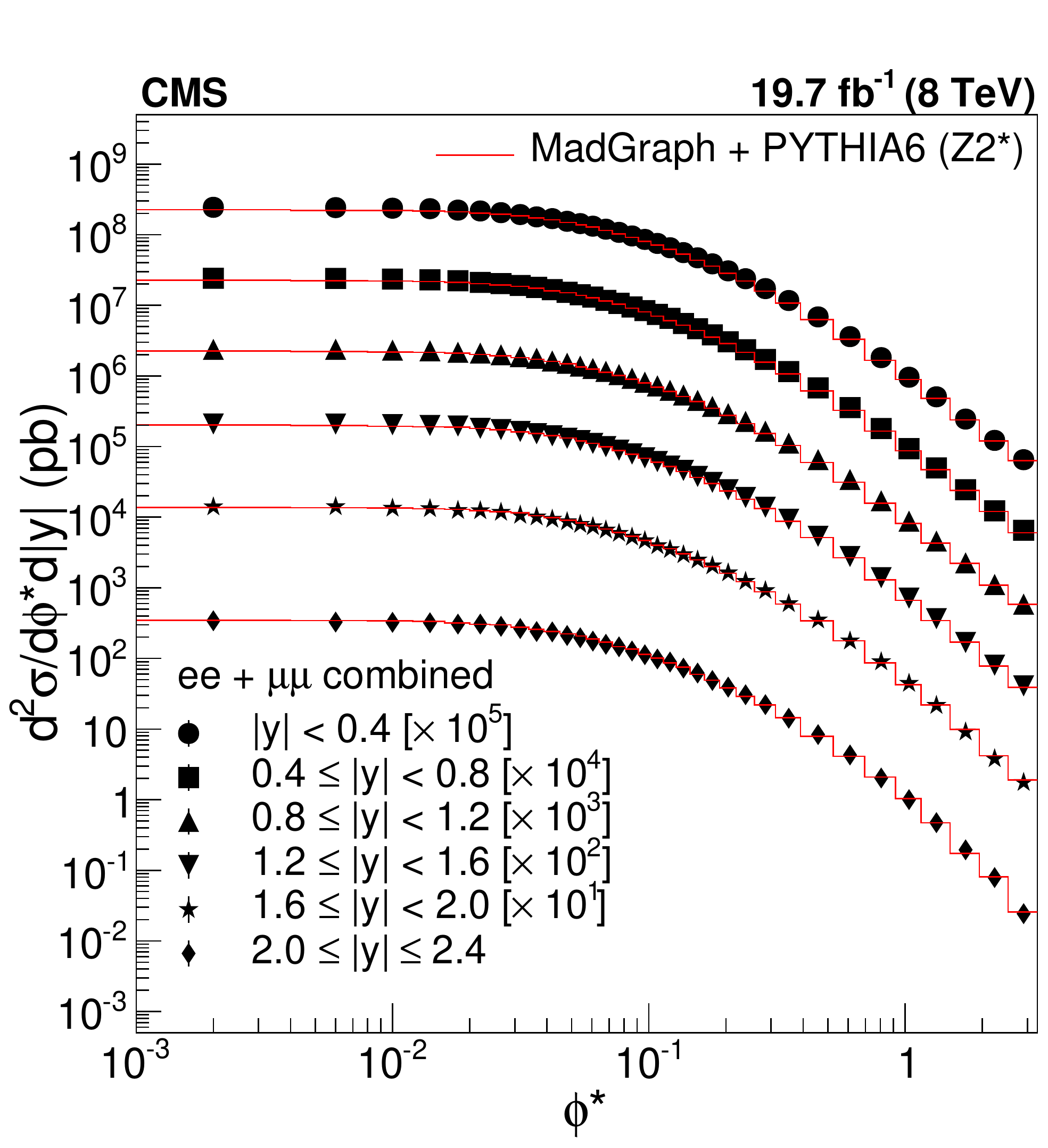}
\includegraphics[width=0.48\textwidth]{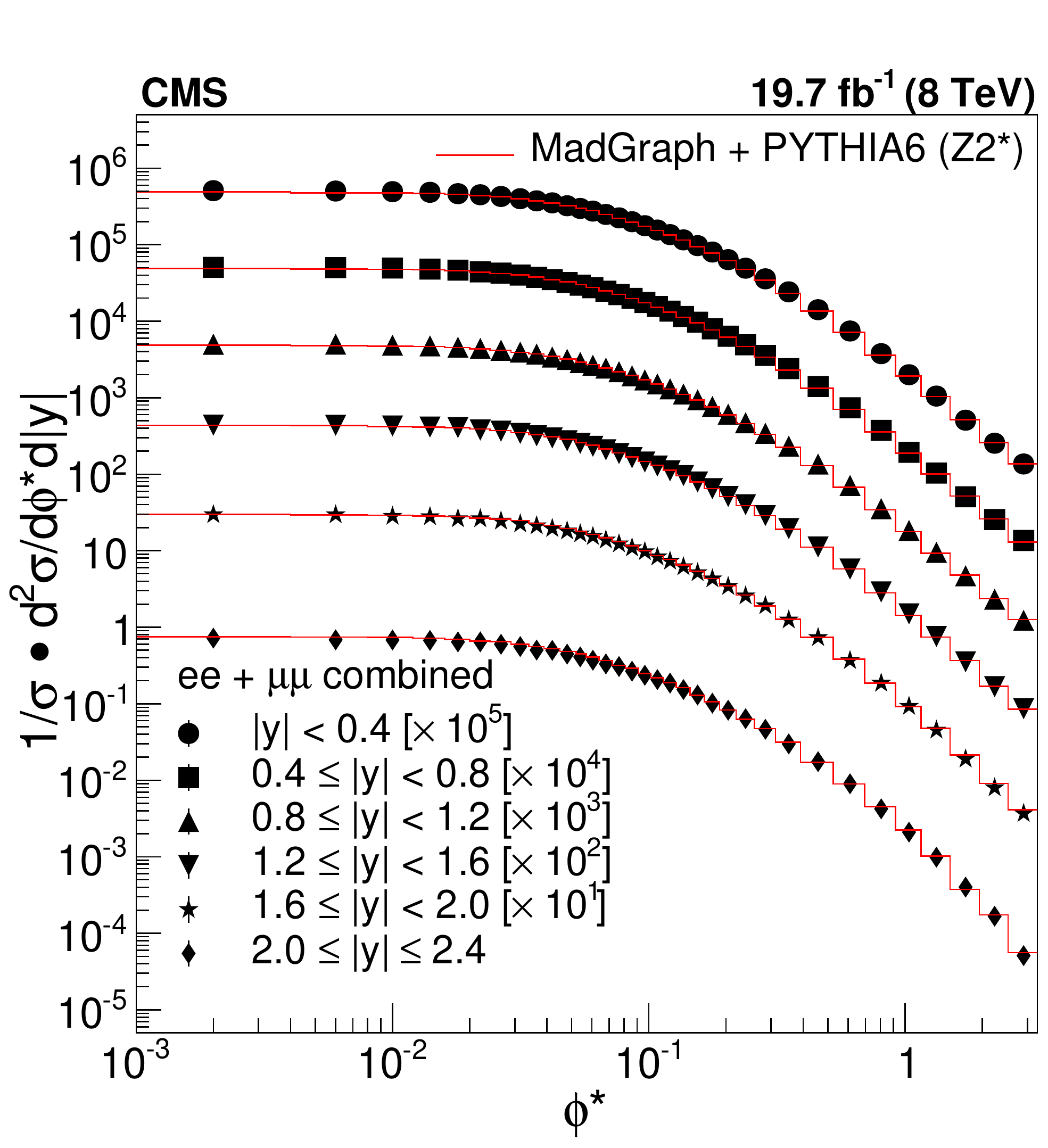}
\caption{The combined absolute (left) and the normalised (right) double-differential cross sections as a function of $\phi^*$ for six ranges of~$\abs{y}$. Experimental data is compared with prediction from  \MADGRAPH{}+\PYTHIASIX with Z2* tune.}
\label{fig:Results2D}
\end{figure}

\begin{figure}[htbp]
\centering
\includegraphics[width=0.98\textwidth]{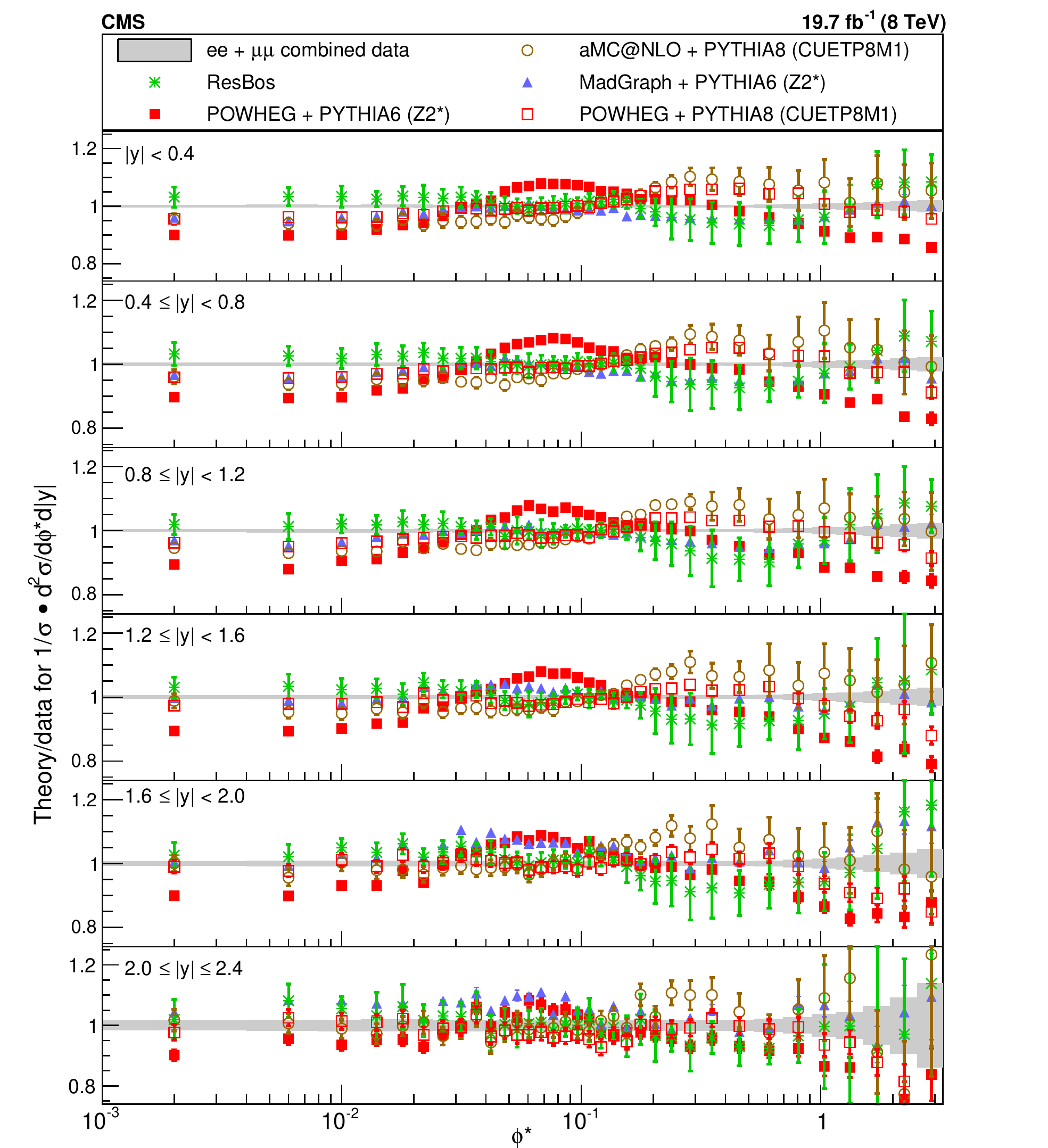}
\caption{The ratio of predicted over measured normalised differential cross sections,
$(1 / \sigma) \, {\rd}^2 \sigma/ {\rd} \phi^* {\rd}\abs{y}$,
  as a function of~$\phi^*$ for six bins in $\abs{y}$.
  The theoretical predictions from \MADGRAPH{}+\PYTHIASIX,
  \POWHEG{}+\PYTHIASIX,   \POWHEG{}+\PYTHIAEIGHT, \RESBOS,
  and \aMCATNLO{}+\PYTHIAEIGHT are shown. The horizontal band corresponds to the uncertainty in the experimental
  measurement. The vertical bars are dominated by the statistical uncertainties in the theoretical predictions.}
\label{fig:AllRatioNorm2D}

\end{figure}

\begin{figure}[htbp]
\centering
\includegraphics[width=0.98\textwidth]{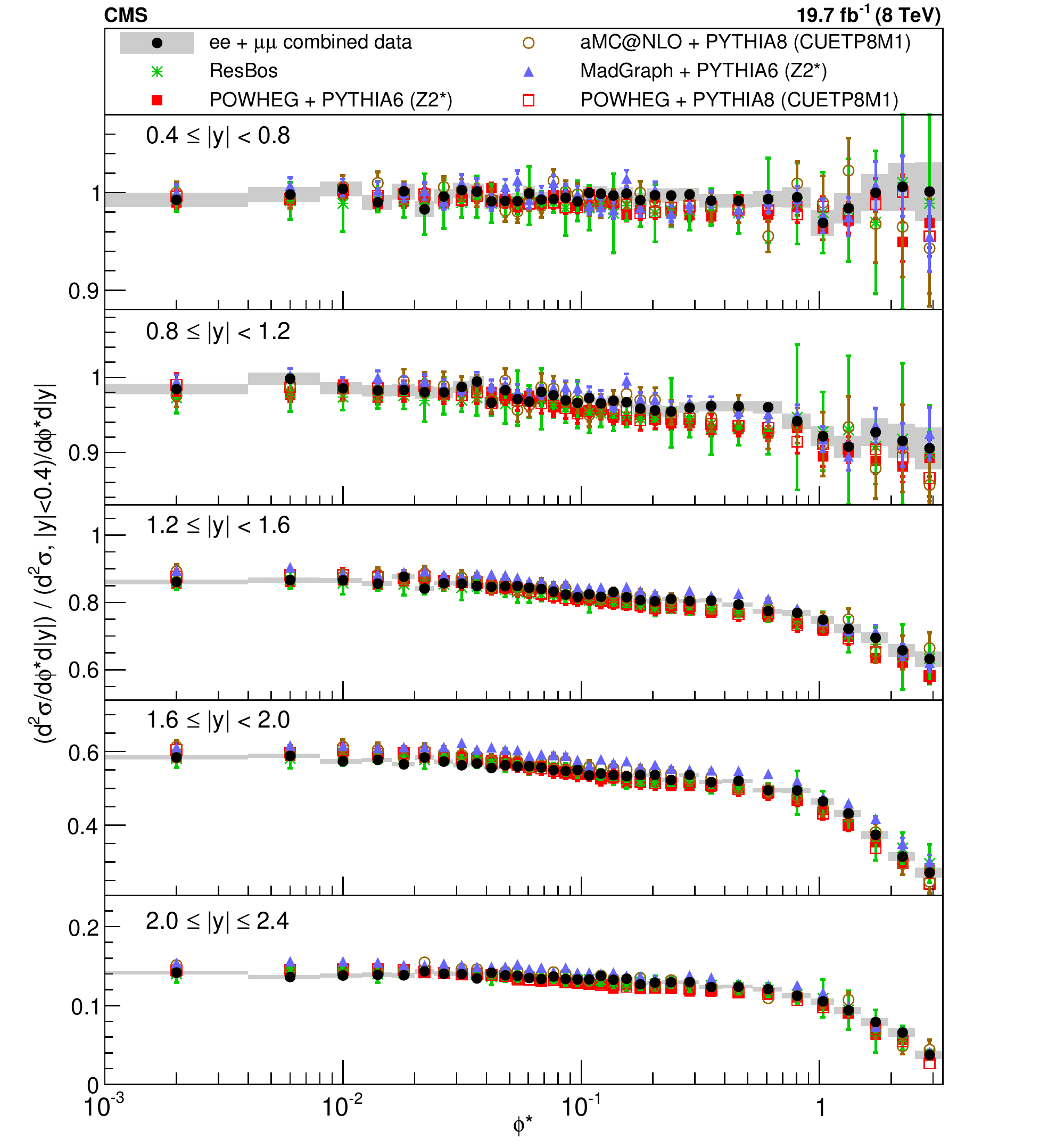}
\caption{The ratio of ${\rd}^2 \sigma / {\rd} \phi^* {\rd}\abs{y}$ for higher rapidity bins
($\abs{y} > 0.4$) normalised to the values in the most central bin
$\abs{y} < 0.4$. The theoretical predictions from \MADGRAPH{}+\PYTHIASIX,
\POWHEG{}+\PYTHIASIX, \POWHEG{}+\PYTHIAEIGHT, \RESBOS, and \aMCATNLO{}+\PYTHIAEIGHT are also shown.
The uncertainties in the theoretical predictions at large $\phi^*$ are dominated by the statistical component.}
\label{fig:ydependence}

\end{figure}

The combined double-differential cross sections are shown in Fig.~\ref{fig:Results2D} with theoretical predictions from \MADGRAPH{}+\PYTHIASIX with Z2* tune. Comparisons with a variety of theoretical predictions for the normalised
cross section are presented in Fig.~\ref{fig:AllRatioNorm2D}. The shape of the $\phi^*$ distribution
varies with dilepton rapidity.  In order to emphasize this feature, ratios of cross sections
as functions of~$\phi^*$ for bins of $\abs{y}$
relative to the central bin $\abs{y} < 0.4$ are presented in Fig.~\ref{fig:ydependence},
where they are compared to predictions
from theoretical calculations and models. All of the theoretical predictions
provide a fairly good description of the shape of the $\phi^*$ distribution with~$\abs{y}$.
However, the predictions from \aMCATNLO{}+\PYTHIAEIGHT and \MADGRAPH{}+\PYTHIASIX  overestimate the cross section at high~$\abs{y}$ by approximately 2\% and 5\%, respectively,
while \POWHEG{}+\PYTHIASIX and \POWHEG{}+\PYTHIAEIGHT underestimate the cross section
by 2\%. The prediction from \RESBOS agrees with the~$\abs{y}$ dependence  at the level of~1\%.

Due to difference in kinematic selections these results cannot be directly compared with similar measurements performed by ATLAS Collaboration~\cite{Aad:2015auj}.

 \section{Summary}
 \label{sec:conclusion}

Measurements of the absolute differential cross sections $\ddinline{\sigma}{\phi^*}$
and ${\rd}^2 \sigma / {\rd} \phi^* {\rd}\abs{y}$ and the corresponding normalised
differential cross sections in the combined dielectron and dimuon channels were  presented for the dilepton mass range of 60 to 120\GeV.
The measurements are based on a sample of proton-proton collision data at a centre-of-mass energy of 8\TeV collected with the CMS detector at the LHC and correspond to an integrated luminosity of 19.7\fbinv. They provide a sensitive test of theoretical predictions.

The normalised cross section $(1/\sigma)\,\ddinline{\sigma}{\phi^*}$
is precise at the level of 0.24--1.2\%.
Theoretical predictions differ from the measurements at the level of 3\%~(\RESBOS), 3\%~(\POWHEG{}+\PYTHIAEIGHT),
 4\%~(\MADGRAPH{}+\PYTHIASIX), 6\%~(\aMCATNLO{}+\PYTHIAEIGHT) and  11\%~(\POWHEG{}+\PYTHIASIX) for $\phi^* \la 0.1$.
For higher values of $\phi^*$ the differences are larger: about 9, 8, 5, 10 and 15\%, respectively.
These observations suggest that more advanced calculations of the
hard-scattering process reproduce the data better.   At the same time,
the large difference in theoretical predictions from a single \POWHEG sample
interfaced with two different versions of \PYTHIA and underlying
event tunes indicates the combined importance of the showering method, nonperturbative effects and the need for soft-gluon resummation on the predicted values of cross sections reported in this paper.

{\tolerance=1200
The variation of the cross section with~$\abs{y}$ is reproduced by \RESBOS within 1\%,
while \MADGRAPH{}+\PYTHIASIX differs from the data by 5\% comparing
the most central and most forward rapidity bins. The predictions from
\aMCATNLO{}+\PYTHIAEIGHT, \POWHEG{}+\PYTHIASIX, and
\POWHEG{}+\PYTHIAEIGHT deviate from the measurement by at most 2\%.
\par}

This analysis validates the overall theoretical description of inclusive production of  vector bosons at the LHC energies by
the perturbative formalism of the standard model.  Nevertheless,
further tuning of the description of the underlying
event is necessary for an accurate prediction of
the kinematics of the Drell--Yan production of lepton pairs.

\begin{acknowledgments}
We congratulate our colleagues in the CERN accelerator departments for the excellent performance of the LHC and thank the technical and administrative staffs at CERN and at other CMS institutes for their contributions to the success of the CMS effort. In addition, we gratefully acknowledge the computing centers and personnel of the Worldwide LHC Computing Grid for delivering so effectively the computing infrastructure essential to our analyses. Finally, we acknowledge the enduring support for the construction and operation of the LHC and the CMS detector provided by the following funding agencies: BMWFW and FWF (Austria); FNRS and FWO (Belgium); CNPq, CAPES, FAPERJ, and FAPESP (Brazil); MES (Bulgaria); CERN; CAS, MoST, and NSFC (China); COLCIENCIAS (Colombia); MSES and CSF (Croatia); RPF (Cyprus); SENESCYT (Ecuador); MoER, ERC IUT, and ERDF (Estonia); Academy of Finland, MEC, and HIP (Finland); CEA and CNRS/IN2P3 (France); BMBF, DFG, and HGF (Germany); GSRT (Greece); OTKA and NIH (Hungary); DAE and DST (India); IPM (Iran); SFI (Ireland); INFN (Italy); MSIP and NRF (Republic of Korea); LAS (Lithuania); MOE and UM (Malaysia); BUAP, CINVESTAV, CONACYT, LNS, SEP, and UASLP-FAI (Mexico); MBIE (New Zealand); PAEC (Pakistan); MSHE and NSC (Poland); FCT (Portugal); JINR (Dubna); MON, RosAtom, RAS, RFBR and RAEP (Russia); MESTD (Serbia); SEIDI, CPAN, PCTI and FEDER (Spain); Swiss Funding Agencies (Switzerland); MST (Taipei); ThEPCenter, IPST, STAR, and NSTDA (Thailand); TUBITAK and TAEK (Turkey); NASU and SFFR (Ukraine); STFC (United Kingdom); DOE and NSF (USA).

\hyphenation{Rachada-pisek} Individuals have received support from the Marie-Curie program and the European Research Council and Horizon 2020 Grant, contract No. 675440 (European Union); the Leventis Foundation; the A. P. Sloan Foundation; the Alexander von Humboldt Foundation; the Belgian Federal Science Policy Office; the Fonds pour la Formation \`a la Recherche dans l'Industrie et dans l'Agriculture (FRIA-Belgium); the Agentschap voor Innovatie door Wetenschap en Technologie (IWT-Belgium); the Ministry of Education, Youth and Sports (MEYS) of the Czech Republic; the Council of Science and Industrial Research, India; the HOMING PLUS program of the Foundation for Polish Science, cofinanced from European Union, Regional Development Fund, the Mobility Plus program of the Ministry of Science and Higher Education, the National Science Center (Poland), contracts Harmonia 2014/14/M/ST2/00428, Opus 2014/13/B/ST2/02543, 2014/15/B/ST2/03998, and 2015/19/B/ST2/02861, Sonata-bis 2012/07/E/ST2/01406; the National Priorities Research Program by Qatar National Research Fund; the Programa Severo Ochoa del Principado de Asturias; the Thalis and Aristeia programs cofinanced by EU-ESF and the Greek NSRF; the Rachadapisek Sompot Fund for Postdoctoral Fellowship, Chulalongkorn University and the Chulalongkorn Academic into Its 2nd Century Project Advancement Project (Thailand); the Welch Foundation, contract C-1845; and the Weston Havens Foundation (USA). \end{acknowledgments}

\bibliography{auto_generated}

\cleardoublepage \appendix\section{The CMS Collaboration \label{app:collab}}\begin{sloppypar}\hyphenpenalty=5000\widowpenalty=500\clubpenalty=5000\textbf{Yerevan Physics Institute,  Yerevan,  Armenia}\\*[0pt]
A.M.~Sirunyan, A.~Tumasyan
\vskip\cmsinstskip
\textbf{Institut f\"{u}r Hochenergiephysik,  Wien,  Austria}\\*[0pt]
W.~Adam, F.~Ambrogi, E.~Asilar, T.~Bergauer, J.~Brandstetter, E.~Brondolin, M.~Dragicevic, J.~Er\"{o}, M.~Flechl, M.~Friedl, R.~Fr\"{u}hwirth\cmsAuthorMark{1}, V.M.~Ghete, J.~Grossmann, J.~Hrubec, M.~Jeitler\cmsAuthorMark{1}, A.~K\"{o}nig, N.~Krammer, I.~Kr\"{a}tschmer, D.~Liko, T.~Madlener, I.~Mikulec, E.~Pree, N.~Rad, H.~Rohringer, J.~Schieck\cmsAuthorMark{1}, R.~Sch\"{o}fbeck, M.~Spanring, D.~Spitzbart, W.~Waltenberger, J.~Wittmann, C.-E.~Wulz\cmsAuthorMark{1}, M.~Zarucki
% ...
\vskip\cmsinstskip
% ...
1:~~Also at Vienna University of Technology, Vienna, Austria\\
% ...

\end{sloppypar}
\end{document}